\documentclass[a4paper,11pt]{article}
\usepackage{pos}

\usepackage{enumitem}
\usepackage{xcolor}
\usepackage{color}
\usepackage{mathtools}
\usepackage{caption} 
\usepackage{subcaption} 
\usepackage{euscript} 
\usepackage{float}
\usepackage{diagbox}
\usepackage[title]{appendix}
\usepackage{tabularx}
\usepackage{tikz}
\usepackage{pgfplots} 
\usepackage{wrapfig}
\pgfplotsset{compat=1.17} 
\usepackage{tcolorbox}
\definecolor{OliveGreen}{rgb}{0,0.6,0}

\usetikzlibrary{decorations.pathmorphing,arrows.meta}

\usepackage{amsmath}
\allowdisplaybreaks
\usepackage{amsfonts}
\usepackage{bbm}
\usepackage{verbatim}
\usepackage{dsfont}
\usepackage{booktabs}
\usepackage{slashed}
\usepackage{braket}

\newcommand{\Le}{{\rm L}}
\newcommand{\Ex}{{\rm E}}
\newcommand{\Ri}{{\rm R}}
\newcommand{\hA}{\widehat{\cal A}}

\renewcommand\d{{\rm d}}

\newcommand{\F}{{\cal F}}
\newcommand{\A}{{\cal A}}

\newcommand{\C}{{\cal C}}

\renewcommand{\S}{{\cal S}}

\newcommand{\be}{\begin{equation}}
\newcommand{\ee}{\end{equation}}

\renewcommand{\thefootnote}{\fnsymbol{footnote}}

\newcommand{\Eq}[1]{Eq.~\eqref{#1}}

\newcommand{\Refs}[1]{Refs~\cite{#1}}
\newcommand{\Sect}[1]{Section~\ref{#1}}
\newcommand{\Sects}[1]{Sections~\ref{#1}}

\newcommand{\Fig}[1]{Figure~\ref{#1}}
\newcommand{\Figs}[1]{Figures~\ref{#1}}

\renewcommand{\O}{{\cal O}}

\newcommand{\ie}{{\it i.e.} }

\newcommand{\eg}{{\it e.g.} }

\renewcommand{\and}{\mbox{and}}

\newcommand{\red}{\color{red}}

\title{Holographic description of closed FRW cosmologies and time-dependent ER=EPR}

\author*[a]{Fran\c cois Rondeau}

\affiliation[a]{Department of Physics, University of Cyprus,\\
  Nicosia 1678, Cyprus}

\emailAdd{rondeau.francois@ucy.ac.cy}

\abstract{We build a covariant holographic entanglement entropy prescription for a class of closed FRW cosmologies, generalizing a recent holographic proposal in de Sitter space. Starting from the Bousso covariant entropy bound, we describe the location of two holographic screens associated with a pair of antipodal observers, and then state our holographic proposal. We then apply our prescription to compute the entanglement entropy of the two-screen and the single screen systems, focusing on the leading classical contributions of order $(G\hbar)^{-1}$. First, we show how the full spacetime is expected to be holographically encoded on the two screens. Second, we argue that the exterior region between the two screens behaves as an Einstein-Rosen bridge, arising from the entanglement between the holographic degrees of freedom as suggested by the ER=EPR conjecture. The entanglement between the two screens, or from the geometric point of view the area of the minimal extremal surface, varies during the cosmological evolution, hence entailing a time-dependent ER=EPR realization. This talk is based on \cite{Franken:2023jas}.}

\FullConference{Corfu Summer Institute 2023 "School and Workshops on Elementary Particle Physics and Gravity" (CORFU2023)\\
 23 April - 6 May , and 27 August - 1 October, 2023\\
Corfu, Greece\\}

\tableofcontents

\begin{document}
\maketitle

\newpage
\section{Introduction}

The notion of holography in High Energy Physics denotes a duality between a theory of quantum gravity in $(n+1)$ spacetime dimensions and a non-gravitational Quantum Field Theory in $n$ spacetime dimensions. This conjecture has been explicitly realized through the gauge/gravity or AdS/CFT correspondence by Maldacena \cite{Maldacena:1997re}, and further developed in \cite{Gubser:1998bc,Witten:1998qj}. One central question in the development of the AdS/CFT correspondence is:
\begin{center} \textbf{\emph{How is the bulk geometric picture related to the dual QFT dynamics?}}
\end{center}

\begin{wrapfigure}{r}{0.36\textwidth}
\centering
	\includegraphics[height=25mm]{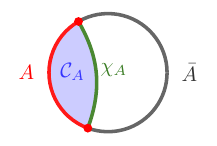}
\caption{\footnotesize Constant time slice of a cylinder where lives a QFT (boundary circle) dual a gravitational theory living in the bulk (interior disk). The Ryu-Takayanagi proposal states that the entanglement entropy between $A$ and $\bar A$ is given by the area of a minimal codimension-$2$ bulk surface $\chi_A$ homologous to $A$. \label{fig:RT_surface}}
\end{wrapfigure}

Following the pioneer work of Ryu-Takayanagi (RT) \cite{Ryu:2006bv,Ryu:2006ef}, the answer to this question turned out to be closely related to the entanglement structure of the dual theory. In a time-independent setup, their proposal asserts that the entanglement entropy of a spatial subsystem $A$ of the dual theory is given by the area of a codimension-$2$ minimal surface $\chi_A$ extended in the bulk: 
\begin{equation}
S(A)=\underset{\chi_A}{\rm min} \frac{{\rm Area (\chi_A})}{4G\hbar},
\end{equation}
where $G$ is the $(n+1)$-dimensional Newton constant, and $\chi_A$ is homologous to $A$, \ie such that there exists a codimension-$1$ surface $\C_A$ satisfying $\partial\C_A=\chi_A\cup A$, see \Fig{fig:RT_surface}.

This prescription has then been generalised by Hubeny-Rangamani-Takayanagi (HRT) to include time dependent situations \cite{Hubeny:2007xt}. Their proposal gives the entanglement entropy of a subregion $A$ of the dual theory in terms of the area of a codimension-$2$ surface $\chi_A$ with \emph{extremal} area, homologous to $A$. If several extremal surfaces exist, we pick among them the one with the minimal area:
\begin{equation}
S(A)={\rm min}~\underset{\chi_A}{\rm ext} \frac{{\rm Area (\chi_A})}{4G\hbar}.
\end{equation}
Such surfaces are thus called \emph{minimal extremal surfaces}, or \emph{HRT surfaces}. Quantum corrections to the HRT proposal have been studied in \cite{Faulkner:2013ana,Engelhardt:2014gca}, while equivalent prescriptions have been found in \cite{Wall:2012uf,Freedman:2016zud,Headrick:2022nbe}. The RT and HRT proposals have been derived in \cite{Lewkowycz:2013nqa} and \cite{Dong:2016hjy} respectively.\\

\begin{figure}[h!]
\hspace{0.5cm}
    \begin{minipage}[c]{.3\linewidth}
        \begin{figure}[H]
        \centering
	\includegraphics[height=25mm]{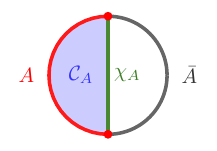}
	\end{figure}
	
    \end{minipage}
	\begin{tikzpicture}
	\draw [-stealth](-1.5,-1) to node[midway, above] {Decreasing $S(A)$}(1.5,-1);
	\end{tikzpicture}
	\hspace{0.3cm}
    \begin{minipage}[c]{.3\linewidth}
    	\begin{figure}[H]
        \centering
	\includegraphics[height=25mm]{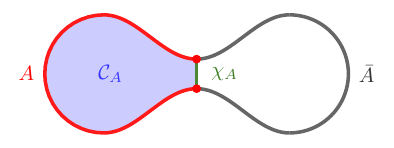}
	\end{figure}
    	
    \end{minipage}
    \caption{\footnotesize Intuitive picture showing the relation between spatial connectivity of the bulk spacetime and quantum entanglement among the holographic degrees of freedom. Decreasing the entanglement entropy $S(A)$ between $A$ and its complement in the dual CFT amounts to pull apart and pinch the associated regions in the bulk geometry.  \label{fig:intuitive_picture}}
\end{figure}

These prescriptions have led to a beautiful interpretation of the connectivity and smoothness of spacetime in terms of quantum entanglement between the holographic degrees of freedom \cite{VanRaamsdonk:2009ar,VanRaamsdonk:2010pw}. An intuitive picture, originally due to Van Raamsdonk, is the following. Let us consider a CFT on S$^n$, and divide the sphere into two hemispheres $A$ and $\bar A$. We denote by $S(A)$ the entanglement entropy between the CFT degrees of freedom associated with the spatial region $A$ and its complement. According to the RT proposal, $S(A)$ is given by the area of a minimal surface $\chi_A$ in the dual spacetime. If one varies the quantum state in such a way that $S(A)$ decreases, so does the area of $\chi_A$: the two spacetime regions separated by $\chi_A$ pull apart and pinch off from each other, see \Fig{fig:intuitive_picture}.
In the limit where $S(A)=0$, the two spacetime regions bounded by $A\cup \chi_A$ and $\bar A \cup \chi_A$ become disconnected from each other. In other words, disentangling degrees of freedom in the dual theory destroys spatial connectivity in the bulk picture.\\

\begin{figure}[h!]
\begin{subfigure}[t]{0.48\linewidth}
\centering
\begin{tikzpicture}
\fill[fill=blue!20] (-1.5-0.4,1.5) -- (-0.4,0) -- (-1.5-0.4,-1.5); 
\draw (-1.5-0.4,1.5) -- (-0.4,0) -- (-1.5-0.4,-1.5) -- node[midway, above, sloped] {CFT$_L$} (-1.5-0.4,1.5);
$\mathbf{\bigotimes}$
\fill[fill=blue!20] (1.5+0.4,1.5) -- (0+0.4,0) -- (1.5+0.4,-1.5); 
\draw (1.5+0.4,-1.5) -- (0+0.4,0) -- (1.5+0.4,1.5) -- node[midway, above, sloped] {CFT$_R$} (1.5+0.4,-1.5);
\end{tikzpicture} 
\caption{\footnotesize The bulk spacetime dual to a product state of the form $\Ket{\psi}=\Ket{\psi}_L\otimes \Ket{\psi}_R$ is a disconnected pair of AdS spacetimes. \label{fig:disconnected_bulk}}
\end{subfigure}\hfill
\begin{subfigure}[t]{0.48\linewidth}
\centering
\begin{tikzpicture}
\fill[fill=blue!20] (-1.5,1.5) -- (0,0) -- (-1.5,-1.5); 
\fill[fill=blue!20] (1.5,1.5) -- (0,0) -- (1.5,-1.5); 
\draw (-1.5,1.5) -- (0,0) -- (-1.5,-1.5) -- node[midway, above, sloped] {CFT$_L$} (-1.5,1.5);
\draw (1.5,-1.5) -- (0,0) -- (1.5,1.5) -- node[midway, above, sloped] {CFT$_R$} (1.5,-1.5);
\draw[decorate,decoration=snake] (-1.5,1.5) -- (1.5,1.5);
\draw[decorate,decoration=snake] (-1.5,-1.5) -- (1.5,-1.5);
\end{tikzpicture}
\caption{\footnotesize The bulk spacetime dual to the thermofield double state is a connected spacetime, with two asymptotic AdS regions (blue shaded triangles) connected by a wormhole (white region). \label{fig:connected_bulk}}
\end{subfigure}\hfill
\caption{\footnotesize Classical spacetimes dual to quantum states in a system of two independent CFTs. \label{fig:}}
\end{figure}
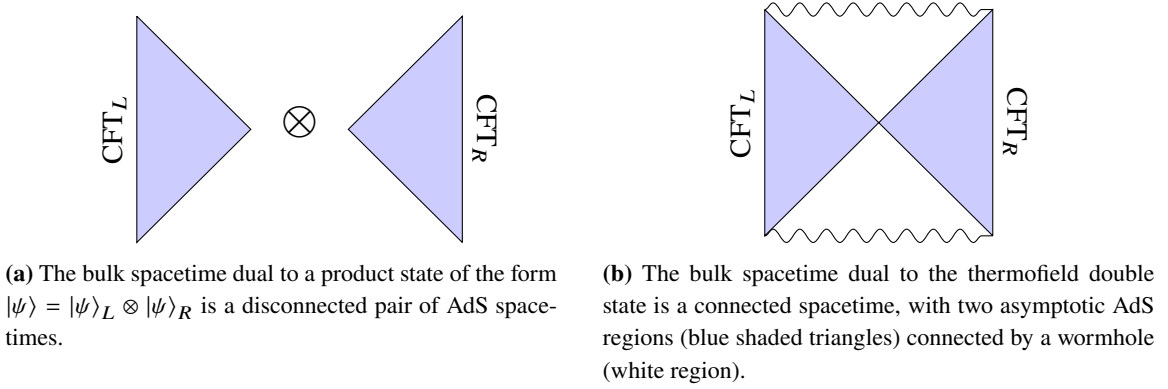

Let us now consider two independent (non-interacting) copies of the CFT, denoted CFT$_L$ and CFT$_R$. The full Hilbert space describing this system is written as $\mathcal{H}=\mathcal{H}_L\otimes\mathcal{H}_R$. Two states of particular interest are:
\begin{itemize}
\item A state with no entanglement between the two systems:
\begin{equation}\label{eq:pure_state}
\Ket{\psi}=\Ket{\psi}_L\otimes \Ket{\psi}_R,
\end{equation}
with $\Ket{\psi}_L\in\mathcal{H}_L$ and $\Ket{\psi}_R\in\mathcal{H}_R$, each of them dual to one asymptotically AdS spacetime. Since the two CFT are independent, the product state \eqref{eq:pure_state} will be dual to a pair of two disconnected independent AdS spacetimes, see \Fig{fig:disconnected_bulk}. 
\item A state with entanglement between the two systems, the so-called \emph{thermofield double state}:
\begin{equation}\label{eq:thermo_field_double_state}
\Ket{\psi_{\rm TFD}}=\frac{1}{\sqrt{Z(\beta)}}\sum_i e^{-\frac{\beta E_i}{2}} \Ket{E_i}_L \otimes \Ket{E_i}_R, 
\end{equation}
where $\Ket{E_i}_{L,R}$ are the energy eigenstates of the two systems, and $Z(\beta)$ the partition function of one CFT at temperature $\beta^{-1}$. This state being a linear combination of product states of the form \eqref{eq:pure_state}, its bulk dual is a quantum superposition of the disconnected spacetimes depicted in \Fig{fig:disconnected_bulk}. On the other hand, Maldacena has argued in \cite{Maldacena:2001kr} that the thermofield double state \eqref{eq:thermo_field_double_state} is dual to the maximally extended AdS-Schwarzschild black hole, whose Penrose diagram is depicted in \Fig{fig:connected_bulk}. This is a classically connected spacetime, with two asymptotic AdS regions connected by a wormhole.
\end{itemize}
We thus arrive at the following conclusion: by entangling the degrees of freedom of the two CFTs, we have glued together the two asymptotic AdS regions in the bulk picture, connecting them through a wormhole (white region in \Fig{fig:connected_bulk}). This idea was further developed under the name ``ER=EPR'' by Maldacena and Susskind \cite{Maldacena:2013xja}, who conjectured that two entangled subsystems (originally discussed by Einstein, Podolsky and Rosen (EPR)) are geometrically connected through an ``Einstein-Rosen'' (ER) bridge, \ie a wormhole. This can be summarized into the formulation ``\emph{entanglement builds spatial bridges}''. Let us note that an algebraic definition of the ER=EPR proposal has recently been proposed in \cite{Engelhardt:2023xer}.\\

\begin{figure}[h!]
    	\centering
\begin{tikzpicture}
\begin{scope}[transparency group]
\begin{scope}[blend mode=multiply]
\path
       +(3,3)  coordinate (IItopright)
       +(-3,3) coordinate (IItopleft)
       +(3,-3) coordinate (IIbotright)
       +(-3,-3) coordinate(IIbotleft)
      
       ;
\draw (IItopleft) --
          node[midway, above, sloped]    {$\cal{J}^+$}
      (IItopright) --
          node[midway, above, sloped] {Antipode}
      (IIbotright) -- 
          node[midway, below, sloped] {$\cal{J}^-$}
      (IIbotleft) --
          node[midway, above , sloped] {Pode}
      (IItopleft) -- cycle;




\fill[fill=blue!20] (-3,3) -- (0,0) -- (-3,-3);
\fill[fill=blue!20] (3,3) -- (0,0) -- (3,-3);

\draw (IItopleft) -- (IIbotright)
              (IItopright) -- (IIbotleft) ;

\node at (-1,1) [circle, fill, inner sep=2 pt, label = below:$\mathcal{S}_1$]{};
\node at (1.5,1.5) [circle, fill, inner sep=2 pt, label = below:$\mathcal{S}_2$]{};



\end{scope}
\end{scope}
\end{tikzpicture}
\caption{\footnotesize Penrose diagram for $(n+1)$-dimensional de Sitter space ($n \geq 2$). Each spacelike slice has the topology of an S$^n$ sphere, and every point represents an S$^{n-1}$, except for points on the left and right vertical edges, which are actual points corresponding to the north and south poles (pode and antipode) of S$^n$ where two antipodal observers are located. The diagonal lines are cosmological horizons for those observers, which bound their static patches (blue shaded regions). The static patch holographic proposal associates to each static patch an holographic screen $\S_1$ and $\S_2$ (black dots), evolving along the cosmological horizons. The exterior region (white region) behaves as a bridge arising from the entanglement between $\S_1$ and $\S_2$.  \label{fig:Penrose_diag_dS}}
\end{figure}
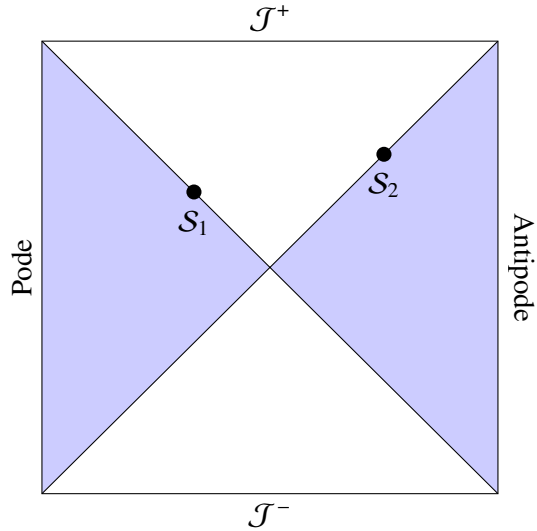

In the famous AdS/CFT correspondence, the dual non-gravitational theory lives at the spatial boundary of the AdS bulk. This naturally raises the question:
\begin{center} \textbf{\emph{Where to locate the dual theory in a spacetime with no spatial boundary?}}
\end{center}
An important example of spacetime with no spatial boundary is the de Sitter (dS) spacetime, the maximally symmetric spacetime with positive curvature, which is believed to describe in good approximation both the very early and current phases of our Universe. The causal structure of dS space is encoded in its Penrose diagram depicted in \Fig{fig:Penrose_diag_dS}.
Any spatial slice of $(n+1)$-dimensional dS has the topology of an S$^{n}$ sphere, which has no spatial boundary. The location of a potential dual quantum mechanical theory is thus much less obvious than in the AdS case, and still debated today. Two main directions have been followed so far:
\begin{itemize}
\item The dS/CFT correspondence \cite{Strominger:2001pn}, where the holographic dual theory is located at future infinity $\cal{J}^+$. See \eg \cite{Strominger:2001gp,Anninos:2011ui,Maldacena:2019cbz,Cotler:2019nbi,Hikida:2021ese,Hikida:2022ltr} for further developments of this correspondence.
\item The static patch holography \cite{Susskind:2021omt,Susskind:2021esx,Shaghoulian:2021cef,Shaghoulian:2022fop}, where the holographic dual theory is located on the cosmological horizon bounding a static patch.
\end{itemize}
The motivation for the de Sitter static patch holographic proposal is the following. While global dS has no spatial boundary, an observer in dS is surrounded by a cosmological horizon which acts as an effective boundary for his/her causally accessible region, \ie his/her static patch. An holographic screen $\S_1$ is then located on this effective boundary associated with this observer. While the degrees of freedom on the screen are sufficient to holographically encode the bulk degrees of freedom of the static patch, they cannot encode the bulk physics of the whole dS. To remedy this, we introduce a second observer at the antipodal point of the spatial sphere, and locate a second holographic screen $\S_2$ on his/her cosmological horizon in order to encode the bulk degrees of freedom in his/her static patch. The crucial point to notice is that now there exists complete bulk Cauchy slices fully contained in the union of the two static patches (see \Fig{fig:Penrose_diag_dS}), and which can therefore be holographically encoded on the union of the two screens. In other words, while each screen independently encode its own static patch, the union of both screens is expected to encode the full de Sitter spacetime, as described in \cite{Franken:2023pni}, a recent covariant extension of the original static patch holographic proposal (see \cite{Franken:2024ruw} for a review). The exterior region behaves as an Einstein-Rosen bridge between the two static patches, arising from the entanglement between $\S_1$ and $\S_2$, as suggested by the ER=EPR proposal.

On the other hand, de Sitter spacetime is a particular case among a more general class of closed FRW cosmologies, with similar topological properties. The aim of this talk, based on \cite{Franken:2023jas}, is to investigate the following question:
\begin{center} \textbf{\emph{How to generalise the de Sitter static patch holographic proposal to arbitrary closed FRW cosmologies?}}
\end{center}

The following \Sects{sect:Bousso_bound} and~\ref{sect:basics_closed_FRW} are reminders of the Bousso covariant entropy bound and some basic aspects of closed FRW cosmologies. The reader familiar with these concepts can skip these sections and proceed directly to \Sect{sect:holo_proposal}, where we present our covariant bilayer holographic proposal for closed FRW cosmologies.

\section{Bousso covariant entropy bound}
\label{sect:Bousso_bound}

The starting point of the de Sitter static patch holography \cite{Susskind:2021omt,Susskind:2021esx,Shaghoulian:2021cef,Shaghoulian:2022fop,Franken:2023pni} and the holographic proposal for closed FRW cosmologies \cite{Franken:2023jas} presented in this talk is the \emph{Bousso covariant entropy bound} \cite{Bousso:1999xy,Bousso:2002ju}, which crucially relies on the concept of light-sheet. Considering an arbitrary codimension-$2$ spatial surface $A$, a light-sheet $L(A)$ of $A$ is a codimension-$1$ surface generated by light rays which begin at $A$, extend orthogonally away from $A$ and are of non-positive expansion. The Bousso bound then states that the coarse-grained entropy (\ie thermodynamic entropy) passing through $L(A)$ is bounded from above by one quarter the area of $A$, in Planck units:
\begin{equation}\label{eq:BCEB}
S_{\rm coarse}(L)\leq \frac{{\rm Area}(A)}{4G\hbar}.
\end{equation}

\begin{figure}[h!]
    \centering
    \begin{subfigure}[t]{0.3\linewidth}
\centering
	\includegraphics[height=60mm]{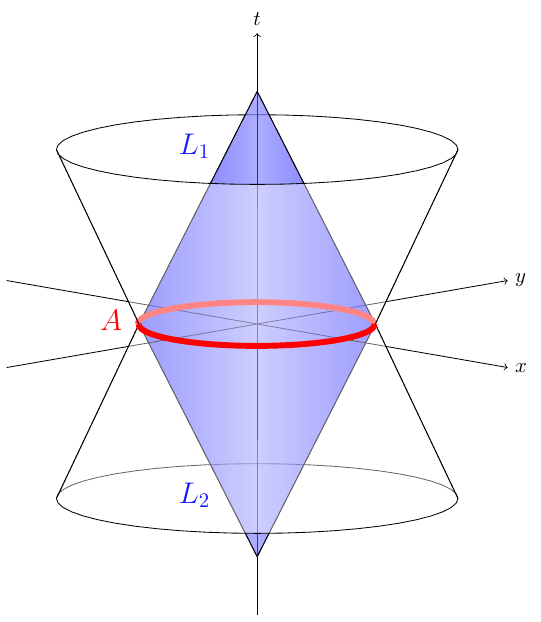}
    \caption{\footnotesize The time direction is the vertical axis, and a codimension-$2$ surface $A$ is depicted by the red circle. Four codimension-$1$ null hypersurfaces emanate orthogonally from $A$: two past directed and two future directed. Two of them have positive expansion (white surfaces), and two have negative expansion (blue shaded surfaces $L_1$ and $L_2$), that are called light-sheets by definition. 
    \label{fig:light_sheets_a}}
    \end{subfigure}\hfill
\begin{subfigure}[t]{0.3\linewidth}
\centering
\begin{tikzpicture}
\path
       +(0,3) coordinate (IItop)
       +(3,0) coordinate (IIright)
       +(0,-3)  coordinate (IIbottom)
      
       ;
\draw (IItop) --
          node[midway, above , sloped] {$\cal{J}^+$}
      (IIright) --
          node[midway, below, sloped] {$\cal{J}^-$}
      (IIbottom) -- 
          node[midway, above , sloped] {} 
       (IItop)--cycle;

\draw[blue] (0,1.5) -- (1.5,0);
\draw[blue] (0,-1.5) -- (1.5,0);
\draw[dashed] (1.5,0) -- (4.5/2,1.5/2);
\draw[dashed] (1.5,0) -- (4.5/2,-1.5/2);


\node at (1.5,0) [circle, fill, inner sep=2 pt,red]{};
\node at (1.5,0) [label=above:$\color{red} A$]{};

\node at (0,1.5) [label=right:$\color{blue} L_1$]{};
\node at (0,-1.5) [label=right:$\color{blue} L_2$]{};

\end{tikzpicture}
\caption{\footnotesize Penrose diagram for $(n+1)$-dimensional Minkowski space $(n\geq 2)$. The codimension-$2$ sphere $A$ is represented by a point, and the two light-sheets $L_1$ and $L_2$ emanating from $A$ by the blue light-like segments. The two codimension-$1$ null hypersurfaces emanating from $A$ with positive expansion are depicted by the dashed light-like segments. \label{fig:light_sheets_b}}
\end{subfigure}\hfill
\begin{subfigure}[t]{0.3\linewidth}
\centering
\begin{tikzpicture}
\path
       +(0,3) coordinate (IItop)
       +(3,0) coordinate (IIright)
       +(0,-3)  coordinate (IIbottom)
      
       ;
\draw (IItop) --
          node[midway, above , sloped] {$\cal{J}^+$}
      (IIright) --
          node[midway, below, sloped] {$\cal{J}^-$}
      (IIbottom) -- 
          node[midway, above , sloped] {} 
       (IItop)--cycle;

\draw (1.3,0.2) -- (1.5,0) -- (1.3,-0.2);


\end{tikzpicture}
\caption{\footnotesize The Bousso wedges in the Penrose diagram indicate the directions of the light-sheets emanating from the codimension-$2$ surface located at the tip of the wedge. \label{fig:light_sheets_c}}
\end{subfigure}\hfill
\caption{\footnotesize Light-sheets associated with a codimension-$2$ sphere $A$ in $(n+1)$-dimensional Minkowski space. The Bousso bound relates the coarse-grained entropy passing through a light-sheet of $A$ to the area of $A$. \label{fig:light_sheets}}
\end{figure}

A codimension-$2$ sphere $A$ is represented by a point on a Penrose diagram, and the four codimension-$1$ null hypersurfaces emanating from $A$ by $45$ degree lines centered on $A$. Among those four directions, the so-called \emph{Bousso wedges} indicate the two directions corresponding to light-sheets. As an illustrative example, the two light-sheets associated with a codimension-$2$ sphere $A$ in $(n+1)$-dimensional ($n\geq 2$) Minkowski space is depicted in \Fig{fig:light_sheets_a}, and on its Penrose diagram in \Fig{fig:light_sheets_b}. The corresponding Bousso wedges are shown on the Penrose diagram of \Fig{fig:light_sheets_c}.

Depending on the orientation of the light-sheets emanating from it, a closed codimension-$2$ surface $A$ can be classified into the three following families:
\begin{itemize}
\item $A$ is called a \emph{normal surface} if it has a future-directed and a past-directed light-sheet on the same side. This is for instance the case for any codimension-$2$ surface in Minkowski space (\Fig{fig:light_sheets}), or in the static patch of an observer in de Sitter space.
\item $A$ is called a \emph{trapped surface} if it has two future-directed light-sheets. This is the case inside a black hole, and in the contracting phase of a cosmological evolution, either in a bouncing universe (\Fig{fig:causal_structure_bouncing}), or in a singular universe close to a Big Crunch singularity (\Fig{fig:causal_structure_singular}).
\item $A$ is called an \emph{anti-trapped surface} if it has two past-directed light-sheets. This is the case in the expanding phase of a cosmological evolution, either in a bouncing universe (\Fig{fig:causal_structure_bouncing}), or in a singular universe close to a Big Bang singularity (\Fig{fig:causal_structure_singular}).
\end{itemize}

The Bousso bound is a classical statement, that has been proven in $4$ dimensions under certain assumptions \cite{Flanagan:1999jp,Bousso:2003kb}, in a hydrodynamic regime where the matter entropy has a phenomenological description in terms of a local entropy current. It is worth mentioning that a formulation of a \emph{Quantum Bousso bound} holding at the semiclassical level has been proposed by Strominger and Thompson \cite{Strominger:2003br}, who conjectured that the area term in the classical Bousso bound should be supplemented by the entanglement entropy across the codimension-$2$ surface $A$. This bound has been proven in \cite{Strominger:2003br} in some vacuum states of the two-dimensional Russo-Susskind-Thorlacius model \cite{Russo:1992ax}, and more recently in any conformal vacua of semiclassical JT gravity \cite{Franken:2023ugu}. For other semiclassical extensions of the Bousso covariant entropy bound, see \cite{Bousso:2014sda,Bousso:2015mna,Bousso:2022tdb}. We will not elaborate further on quantum versions of the Bousso bound, and will stay at the classical level in the following.

\section{Basics of closed FRW cosmology}
\label{sect:basics_closed_FRW}
\begin{figure}[h!]
    \centering
\begin{subfigure}[t]{0.45\linewidth}
\centering
\begin{tikzpicture}

\path
       +(1,5/2) coordinate (IItopright)
       +(-1,5/2) coordinate (IItopleft)
       +(1,-5/2) coordinate (IIbotright)
       +(-1,-5/2) coordinate(IIbotleft)
      
       ;
       
              
\draw (IItopleft) --
          node[midway, above, sloped]    {$\cal{J}^+$}
      (IItopright) --
          node[midway, above, sloped] {Antipode}
      (IIbotright) -- 
          node[midway, below, sloped] {$\cal{J}^-$}
      (IIbotleft) --
          node[midway, above , sloped] {Pode}
      (IItopleft) -- cycle;

\draw (IItopleft) -- (IIbotright)
        (IItopright) -- (IIbotleft) ;



\node at (-1,5/2) [label = left:$|\gamma|\pi$]{};
\node at (-1,-5/2) [label = left:$0$]{};

\node at (-1,-5/2) [label = below:$0$]{};
\node at (1,-5/2) [label = below:$\pi$]{};
	
\end{tikzpicture}   
\caption{\centering \footnotesize Bouncing closed FRW cosmology with $\gamma<-1$.}
\end{subfigure}
\quad \,
\begin{subfigure}[t]{0.45\linewidth}
\centering
\begin{tikzpicture}

\path
       +(1,5/2) coordinate (IItopright)
       +(-1,5/2) coordinate (IItopleft)
       +(1,-5/2) coordinate (IIbotright)
       +(-1,-5/2) coordinate(IIbotleft)
      
       ;
       
              
\draw[decorate,decoration=snake] (IItopleft) --
          node[midway, above, sloped]    {$\phantom{\cal{J}^+}$}
      (IItopright);
      
\draw (IItopright) --
          node[midway, above, sloped] {Antipode}
      (IIbotright);
      
\draw[decorate,decoration=snake]  (IIbotright) -- 
          node[midway, below, sloped] {$\phantom{\cal{J}^-}$}
      (IIbotleft);
      
\draw (IIbotleft) --
          node[midway, above , sloped] {Pode}
      (IItopleft);
      
\draw (IItopleft) -- (IIbotright)
        (IItopright) -- (IIbotleft) ;
        



\node at (-1,5/2) [label = left:$\phantom{||}\gamma\pi$]{};
\node at (-1,-5/2) [label = left:$0$]{};

\node at (-1,-5/2) [label = below:$0$]{};
\node at (1,-5/2) [label = below:$\pi$]{};
	
\end{tikzpicture}     
\caption{\centering \footnotesize Singular closed FRW cosmology with $\gamma>1$.}
\end{subfigure}
   \caption{\footnotesize Penrose diagram for $(n+1)$-dimensional closed FRW cosmologies ($n \geq 2$). Each spacelike slice has the topology of an S$^n$ sphere, and every point represents an S$^{n-1}$, except for points on the left and right vertical edges, which are actual points corresponding to the north and south poles (pode and antipode) of S$^n$. Coordinates are chosen such that the worldlines of two antipodal comoving observers follow the right and left edges of the diagrams. The wavy lines indicate the Big Bang/Big Crunch singularities in the cosmologies with $\gamma\geq 1$. \label{fig:Penrose_diag}}
\end{figure}
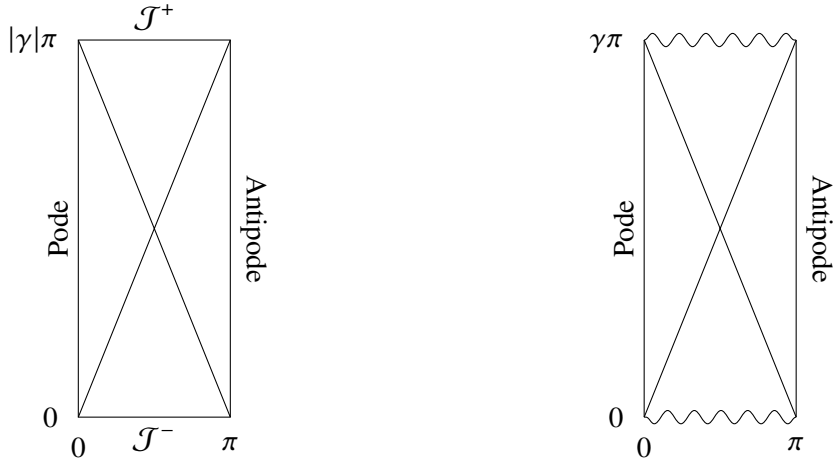

The aim of this section is to remind some well-known features of closed FRW cosmologies, which will be useful for the holographic description of such spacetimes described in the next sections. The metric of an $(n+1)$-dimensional closed FRW cosmology is given by:
\begin{equation}\label{eq:metric}
\d s^2=a^2(\eta)\left[-\d\eta^2+\d\theta^2+\sin^2(\theta)\d\Omega_{n-1}^2\right],
\end{equation}
where $\eta$ is the conformal time, $\theta\in[0,\pi]$ a polar angle, $a(\eta)$ the scale factor, and $\d\Omega_{n-1}^2$ the metric of the sphere S$^{n-1}$ of radius $1$. Denoting respectively by $\rho$ and $p$ the energy density and pressure in the universe, the Friedmann equations read:
\begin{eqnarray}
{n(n-1)\over 2} \left[\Big({a'\over a}\Big)^2+1\right] &=& 8\pi G\, a^2\rho,\\
\rho'+n\, {a'\over a}\, (\rho+p) &=& 0,
\end{eqnarray}
where primes denote derivative with respect to conformal time $\eta$. From now we will assume that the cosmological evolution is induced by a single perfect fluid satisfying the equation of state 
\begin{equation}
p=w \rho,
\end{equation}
where the constant $w\in [-1,1]$ is the perfect fluid index. One can then solve the Friedmann equations to get the explicit evolution of the scale factor:
\begin{equation}\label{eq:scale_factor}
a(\eta)=a_0\!\left(\sin{\eta\over |\gamma|}\right)^\gamma, ~~\quad \eta\in\big[0,|\gamma|\pi\big], 
\end{equation}
where $\gamma\in (-\infty,-1]\cup[1/(n-1),+\infty)$ is a constant related to the perfect fluid index $w$ and the number $n$ of spatial dimensions by:
\begin{equation}
\gamma=\frac{2}{-2+n(w+1)}.
\end{equation}
The cosmological evolution crucially depends on the value of the fluid index $w$, or equivalently $\gamma$:
\begin{itemize}
\item For $\gamma\in (-\infty,-1]$, the scale factor decreases from an infinite value at $\eta=0$, reaches its minimum $a_0$ at $\eta=|\gamma|\pi/2$, and then expands and becomes infinite at finite conformal time $\eta=|\gamma|\pi$. The cosmological evolution bounces and is nowhere singular.
\item For $\gamma\in [1/(n-1),+\infty)$, the scale factor increases from a Big Bang singularity at $\eta=0$ where it vanishes, reaches its maximum $a_0$ at $\eta=|\gamma|\pi/2$, and then decreases up to a Big Crunch singularity at $\eta=|\gamma|\pi$, where it vanishes.
\end{itemize}
The Penrose diagrams describing the causal structure of these cosmologies are rectangles parameterized by the conformal coordinates $(\theta,\eta)$. While the width is constant and equal to $\pi$, the height varies according to $|\gamma|$. For $\gamma \in [1/(n-1),1)$, the Penrose diagram is wider than tall. The bilayer holographic proposal presented in \Sect{sect:holo_proposal} is not expected to apply for this class of cosmologies, as discussed in \cite{Franken:2023jas}. We will not consider this class of cosmologies in the following, and will focus on the cases $|\gamma|\geq 1$, for which the Penrose diagrams are taller than wide, as shown in \Fig{fig:Penrose_diag}.

Some relevant values of $\gamma$ are:
\begin{itemize}
\item $\gamma=-1$, corresponding to a cosmological evolution induced by a positive
cosmological constant, namely de Sitter space.
\item $\gamma=1/(n-1)$, corresponding to a
cosmological evolution induced by moduli fields.
\item $\gamma=2/(n-1)$, corresponding to a cosmological evolution induced by radiation.
\item $\gamma=2/(n-2),~n \geq 3$, corresponding to
a cosmological evolution induced by massive non-relativistic matter.
\end{itemize}

\begin{figure}[h!]
\begin{subfigure}[t]{0.45\linewidth}
\centering
\begin{tikzpicture}

\path
       +(1,5/2) coordinate (IItopright)
       +(-1,5/2) coordinate (IItopleft)
       +(1,-5/2) coordinate (IIbotright)
       +(-1,-5/2) coordinate(IIbotleft)
      
       ;
       
       \fill[fill=blue!20] (-1,5/2) -- (1,1/2) -- (1,-1/2) -- (-1,-5/2) -- cycle;
              
\draw (IItopleft) --
          node[midway, above, sloped]    {}
      (IItopright) --
          node[midway, above, sloped] {}
      (IIbotright) -- 
          node[midway, below, sloped] {}
      (IIbotleft) --
          node[midway, above , sloped] {}
      (IItopleft) -- cycle;

\draw (IItopleft) -- (IIbotright)
        (IItopright) -- (IIbotleft) ;
\draw[dashed] (IItopleft) -- (1,1/2)
                (IItopright) -- (-1,1/2)
                (IIbotright) -- (-1,-1/2)
                (IIbotleft) -- (1,-1/2);



\node at (-1,5/2) [label = left:$|\gamma|\pi$]{};
\node at (-1,-5/2) [label = left:$0$]{};

\node at (-1,-5/2) [label = below:$0$]{};
\node at (1,-5/2) [label = below:$\pi$]{};
	
\end{tikzpicture} 
\begin{tikzpicture}

\path
       +(1,5/2) coordinate (IItopright)
       +(-1,5/2) coordinate (IItopleft)
       +(1,-5/2) coordinate (IIbotright)
       +(-1,-5/2) coordinate(IIbotleft)
      
       ;
       
       \fill[fill=red!20] (IIbotleft) -- (0,0) -- (IIbotright) -- cycle;
       \fill[fill=green!20] (IItopleft) -- (0,0) -- (IItopright) -- cycle;
       
\draw (IItopleft) --
          node[midway, above, sloped]    {}
      (IItopright) --
          node[midway, above, sloped] {}
      (IIbotright) -- 
          node[midway, below, sloped] {}
      (IIbotleft) --
          node[midway, above , sloped] {}
      (IItopleft) -- cycle;

\draw (IItopleft) -- (IIbotright)
        (IItopright) -- (IIbotleft) ;
\draw[dashed] (IItopleft) -- (1,1/2)
                (IItopright) -- (-1,1/2)
                (IIbotright) -- (-1,-1/2)
                (IIbotleft) -- (1,-1/2);


\draw (-0.2,3/2+1/2-1) -- (0,3/2+1/2+0.2-1) -- (0.2,3/2+1/2-1);
\draw (-1/2-0.2,-0.2) -- (-1/2,0) -- (-1/2-0.2,0.2);
\draw (1/2+0.2,-0.2) -- (1/2,0) -- (1/2+0.2,0.2);
\draw (-0.2,-3/2-1/2+1) -- (0,-3/2-1/2-0.2+1) -- (0.2,-3/2-1/2+1);

\node at (-1,5/2) [label = left:$|\gamma|\pi$]{};
\node at (-1,-5/2) [label = left:$0$]{};

\node at (-1,-5/2) [label = below:$0$]{};
\node at (1,-5/2) [label = below:$\pi$]{};
	
\end{tikzpicture} 
\caption{\centering \footnotesize Bouncing closed FRW cosmology with $\gamma<-1$. \label{fig:causal_structure_bouncing}}
\end{subfigure}
\quad \,
\begin{subfigure}[t]{0.45\linewidth}
\centering
\begin{tikzpicture}
\path
       +(1,5/2) coordinate (IItopright)
       +(-1,5/2) coordinate (IItopleft)
       +(1,-5/2) coordinate (IIbotright)
       +(-1,-5/2) coordinate(IIbotleft)
      
       ;
       
       \fill[fill=blue!20] (-1,5/2) -- (1,1/2) -- (1,-1/2) -- (-1,-5/2) -- cycle;
       
\draw[decorate,decoration=snake] (IItopleft) --
          node[midway, above, sloped]    {}
      (IItopright);
      
\draw (IItopright) --
          node[midway, above, sloped] {}
      (IIbotright);
      
\draw[decorate,decoration=snake]  (IIbotright) -- 
          node[midway, below, sloped] {}
      (IIbotleft);
      
\draw (IIbotleft) --
          node[midway, above , sloped] {}
      (IItopleft);
      
\draw (IItopleft) -- (IIbotright)
        (IItopright) -- (IIbotleft) ;



\draw[dashed] (IItopleft) -- (1,1/2)
                (IItopright) -- (-1,1/2)
                (IIbotright) -- (-1,-1/2)
                (IIbotleft) -- (1,-1/2);

\node at (-1,5/2) [label = left:$\phantom{||}\gamma\pi$]{};
\node at (-1,-5/2) [label = left:$0$]{};

\node at (-1,-5/2) [label = below:$0$]{};
\node at (1,-5/2) [label = below:$\pi$]{};
\end{tikzpicture}    
\begin{tikzpicture}

\path
       +(1,5/2) coordinate (IItopright)
       +(-1,5/2) coordinate (IItopleft)
       +(1,-5/2) coordinate (IIbotright)
       +(-1,-5/2) coordinate(IIbotleft)
      
       ;
       
       \fill[fill=green!20] (IIbotright) decorate[decoration=snake] {to (IIbotleft)} -- (0,0) -- cycle;
       \fill[fill=red!20] (IItopleft) decorate[decoration=snake] {to (IItopright)} -- (0,0) -- cycle;
       
\draw[decorate,decoration=snake] (IItopleft) --
          node[midway, above, sloped]    {}
      (IItopright);
      
\draw (IItopright) --
          node[midway, above, sloped] {}
      (IIbotright);
      
\draw[decorate,decoration=snake]  (IIbotright) -- 
          node[midway, below, sloped] {}
      (IIbotleft);
      
\draw (IIbotleft) --
          node[midway, above , sloped] {}
      (IItopleft);
      
\draw (IItopleft) -- (IIbotright)
        (IItopright) -- (IIbotleft) ;
        
\draw[dashed] (IItopleft) -- (1,1/2)
                (IItopright) -- (-1,1/2)
                (IIbotright) -- (-1,-1/2)
                (IIbotleft) -- (1,-1/2);


\draw (-0.2,3/2+1/2+0.2-1.2) -- (0,3/2+1/2-1.2) -- (0.2,3/2+1/2+0.2-1.2);
\draw (-1/2-0.2,-0.2) -- (-1/2,0) -- (-1/2-0.2,0.2);
\draw (1/2+0.2,-0.2) -- (1/2,0) -- (1/2+0.2,0.2);
\draw (-0.2,-3/2-1/2-0.2+1.2) -- (0,-3/2-1/2+1.2) -- (0.2,-3/2-1/2-0.2+1.2);

\node at (-1,5/2) [label = left:$\phantom{||}\gamma\pi$]{};
\node at (-1,-5/2) [label = left:$0$]{};

\node at (-1,-5/2) [label = below:$0$]{};
\node at (1,-5/2) [label = below:$\pi$]{};
	
\end{tikzpicture} 
\caption{\centering \footnotesize Singular closed FRW cosmology with $\gamma>1$. \label{fig:causal_structure_singular}}
\end{subfigure}
   \caption{\footnotesize Causal structure associated with a pair of comoving observers sitting at the pode and the antipode, depicted by the left and right vertical edges of the Penrose diagrams. The cosmological horizons are depicted by the dashed lines, which delimit the causal patches of each observer. The causal patch of the observer sitting at the pode corresponds to the blue shaded region. The apparent horizons are the diagonal lines, which split the Penrose diagrams in four triangular regions where the Bousso wedges are indicated. They are defined as the boundaries of the union of all trapped (red shaded region) and anti-trapped (green shaded region) surfaces.  \label{fig:causal_structure}}
\end{figure}
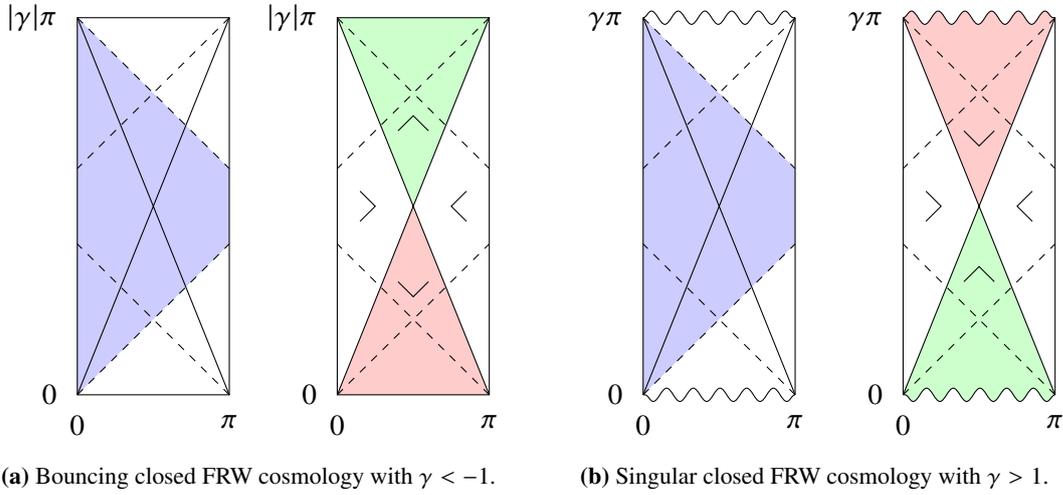

Three different kind of horizons can be associated to a pair of comoving observers located at the pode and antipode:
\begin{itemize}
\item The \emph{particle horizon} delimits the part of spacetime that the observer will ever be able to send a signal to. \item The \emph{event horizon} delimits the part of spacetime that the observer will ever be able to receive a signal from.
\item The \emph{apparent horizons} are the boundaries of the regions of trapped and anti-trapped surfaces.
\end{itemize}
The particle and event horizons, also called cosmological horizons, bound the causal patches of each observer. The causal structure of bouncing and singular closed FRW cosmologies is shown on the Penrose diagrams of \Fig{fig:causal_structure}.
Let us note that the cosmological horizons coincide with the apparent horizons only for $|\gamma|=1$ (and in particular in the dS case $\gamma=-1$), in which case the Penrose diagram is a square.

\section{Closed FRW cosmology holographic proposal}
\label{sect:holo_proposal}

In this section, we motivate our proposal for generalizing the de Sitter static patch holography conjecture~\cite{Susskind:2021omt,Susskind:2021esx,Shaghoulian:2021cef,Shaghoulian:2022fop,Franken:2023pni} to closed FRW cosmologies, when $|\gamma|\ge 1$. We will reach our conclusions by considering an arbitrary foliation $\F$ of spacetime, with SO$(n)$-symmetric Cauchy slices. Denoting by $\Sigma$ a generic slice of $\F$, we will assume that the state on $\Sigma$ is pure, and thus with vanishing entanglement entropy. Our proposal, which relies on the Bousso covariant entropy conjecture~\cite{Bousso:1999xy,Bousso:2002ju} as well as the bilayer prescription of \Refs{Shaghoulian:2021cef,
Shaghoulian:2022fop,Franken:2023pni}, will then be tested in \Sect{sect:time_dep_ER=EPR}. For other related works studying holography and cosmology, see \eg \cite{Fischler:1998st,Hellerman:2001yi,Bak:1999hd,Bousso:1999cb,Diaz:2007mh,Bousso:2015mqa,Bousso:2015qqa,Sanches:2016sxy,Caginalp:2019fyt}. 

\subsection{Location of the dual holographic system}
\label{sect:location_screens}

Let us pick a Cauchy slice $\Sigma$ in $\F$ and consider placing a holographic $(n-1)$-dimensional spherical screen on its part lying in the causal patch of the pode. We will denote this screen $\S_1$. According to the holographic principle \cite{tHooft:1993dmi,Susskind:1994vu}, the area of the screen in Planck units is a measure of the number of the holographic degrees of freedom on the screen. In particular, we would like $\S_1$ to be large enough so as to encode the state on $\Sigma_1$, the part of $\Sigma$ located to the left of the screen and lying entirely in the causal patch of the pode. The screen $\S_1$ must be kept in the causal patch of the observer at the pode in order for the information associated with it to be accessible to this observer. The precise location of $\S_1$ on $\Sigma$ will be determined so as to maximize as much as possible the extent of $\Sigma_1$, which is to be described holographically in accordance with the covariant entropy conjecture~\cite{Bousso:1999xy,Bousso:2002ju}.

\begin{figure}[H]
    \begin{minipage}[c]{.55\linewidth}

Let us initially place the screen $\S_1$ at the intersection of $\Sigma$ with the apparent horizon of the pode, as depicted in \Fig{fig:location_screen_1}. From the orientation of the Bousso wedge in the left triangular region of the Penrose diagram, one sees that $\S_1$ has a future-directed light-sheet $L$, drawn in blue in \Fig{fig:location_screen_1}. On one hand, the Bousso covariant entropy bound \eqref{eq:BCEB} states that $S_{\rm coarse}(L)\leq {\rm Area}(\S_1)/(4G\hbar)$. On the other hand, the second law of thermodynamics ensures that $S_{\rm coarse}(\Sigma_1)\leq S_{\rm coarse}(L)$, so that
\begin{equation}
    S_{\rm coarse}(\Sigma_1)\leq \frac{{\rm Area}(\S_1)}{4G\hbar}.
\end{equation}
Therefore, the sphere $\S_1$ has enough degrees of freedom to holographically encode the state on $\Sigma_1$.
        
    \end{minipage}
    \hfill%
    \begin{minipage}[c]{0.4\linewidth}
	\centering
	\begin{tikzpicture}

\path
       +(2,4)  coordinate (IItopright)
       +(-2,4) coordinate (IItopleft)
       +(2,-4) coordinate (IIbotright)
       +(-2,-4) coordinate(IIbotleft)
      
       ;
       
\draw (IItopleft) --
          node[midway, above, sloped]    {}
      (IItopright) --
          node[midway, above, sloped] {}
      (IIbotright) -- 
          node[midway, below, sloped] {}
      (IIbotleft) --
          node[midway, above , sloped] {}
      (IItopleft) -- cycle;
      
\draw (IItopleft) -- (IIbotright)
              (IItopright) -- (IIbotleft) ;


\draw (-0.2,2.8) -- (0,3) -- (0.2,2.8);
\draw (-1-0.2,-0.2) -- (-1,0) -- (-1-0.2,0.2);
\draw (1+0.2,-0.2) -- (1,0) -- (1+0.2,0.2);
              
\draw[domain=-2:-0.75, smooth, variable=\x, red,line width=0.8mm] plot ({\x}, {sin(deg((\x/2-1)))+2.5});
\draw[domain=-0.75:2, smooth, variable=\x, red] plot ({\x}, {sin(deg((\x/2-1)))+2.5});

\draw[blue] (-0.75,1.53) -- (-2,2.78) ;

\node at (-0.75,1.53) [circle, fill, inner sep=2 pt]{};
\node at (-0.5,1.4) [label=above:$\S_1$]{};

\node at (-1.4,1.6) [label=below:$\color{red} \Sigma_1$]{};
\node at (1.5,2.3) [label=below:$\color{red} \Sigma$]{};

\node at (-2.1,2.9) [label=right:$\color{blue} L$]{};

\end{tikzpicture}
\caption{\label{fig:location_screen_1}}
    \end{minipage}
\end{figure}

One may wonder whether we can push the screen $\S_1$ farther towards the right along $\Sigma$, in order to describe holographically a greater part of it. This depends on the phase of the cosmology where $\Sigma$ sits. Because $\Sigma$ is a spacelike slice, its part located between the apparent horizons lies entirely either in the top or bottom triangular region of the Penrose diagram. It is either in the region of expansion or contraction of the cosmology.

\paragraph{Expanding phase.} Let us start with the situation when the part of $\Sigma$ between the apparent horizons lies in the expanding phase of the cosmology. In this case, the orientation of the Bousso wedges in the expanding phase shows that any codimension-$2$ surface in this region has two past directed light-sheet, but no future directed ones (see \Fig{fig:location_screen_1}). Therefore, the argument presented above to bound the coarse-grained entropy passing through $\Sigma_1$ by the area of $\S_1$ cannot be applied. In the expanding phase of the cosmology, we thus keep $\S_1$ on the apparent horizon of the pode.

\begin{figure}[H]
    \begin{minipage}[c]{.55\linewidth}
    
\paragraph{Contracting phase.} The situation when the part of $\Sigma$ between the apparent horizons lies in the contracting phase of the cosmology is more involved. The orientation of the Bousso wedge in the bottom triangular region shows that any codimension-$2$ surface in this region has a future directed incoming light-sheet, shown in blue in \Fig{fig:location_contracting_phase}. Let us define $\eta_2$ the conformal time at which $\Sigma$ intersects the apparent horizon of the antipode, and $\eta_{\rm c}$ the conformal time at which the particle horizon of the pode intersects the apparent horizon of the antipode, depicted by the red dot in \Fig{fig:location_contracting_phase}.
        
\begin{itemize}
    \item If $\eta_2=\tilde\eta_2\in[0,\eta_{\rm c}]$, $\S_1$ can be pushed at most up to the particle horizon of the pode, in order for the screen and the whole $\Sigma_1$ to stay in the causal patch of the pode.
    \item If $\eta_2=\hat\eta_2\in[\eta_{\rm c},|\gamma|\pi/2]$, $\S_1$ can be pushed up to the apparent horizon of the antipode.
\end{itemize}
       
    \end{minipage}
    \hfill%
    \begin{minipage}[c]{0.4\linewidth}
	\centering
	\begin{tikzpicture}

\path
       +(2,4)  coordinate (IItopright)
       +(-2,4) coordinate (IItopleft)
       +(2,-4) coordinate (IIbotright)
       +(-2,-4) coordinate(IIbotleft)
      
       ;

\fill[fill=violet!20] (-2,-4) -- (2/3,-4/3) -- (0,0) -- cycle;       
       
\draw (IItopleft) --
          node[midway, above, sloped]    {}
      (IItopright) --
          node[midway, above, sloped] {}
      (IIbotright) -- 
          node[midway, below, sloped] {}
      (IIbotleft) --
          node[midway, above , sloped] {}
      (IItopleft) -- cycle;
      
\draw[violet,line width=0.8mm] (IItopleft) -- (0,0);
\draw (0,0) -- (IIbotright);
\draw (IItopright) -- (IIbotleft) ;

\draw[dotted] (2/3,-4/3)--(2,-4/3);   
\node at (2,-4/3) [label = right:$\eta_{\rm c}$]{};

\draw[dotted] (1,-2)--(2,-2);  
\node at (2,-2) [label = right:$\tilde\eta_2$]{};

\draw[dotted] (0.5,-0.84)--(2,-0.84);  
\node at (2,-0.84) [label = right:$\hat\eta_2$]{};

\draw (-0.2,-2.8-0.5) -- (0,-3-0.5) -- (0.2,-2.8-0.5);
\draw (-1-0.2+0.4,-0.2) -- (-1+0.4,0) -- (-1-0.2+0.4,0.2);
\draw (1+0.2-0.4,-0.2) -- (1-0.4,0) -- (1+0.2-0.4,0.2);

\node at (IItopright) [label = right:$|\gamma|\pi$]{};
\node at (IIbotright) [label = right:$0$]{};
              
\draw[dashed,gray] (IIbotleft) -- (2/3,-4/3);


\draw[violet,line width=0.8mm] (-2,-4) to [bend right=21] (0,0) ;

\draw[domain=-2:-0.15, smooth, variable=\x, red,line width=0.8mm] plot ({\x}, {sin(deg((\x/2-1)))-0.1});
\draw[domain=-0.15:2, smooth, variable=\x, red] plot ({\x}, {sin(deg((\x/2-1)))-0.1});

\draw[domain=-2:-0.74, smooth, variable=\x, red,line width=0.8mm] plot ({\x}, {sin(deg((\x/2-1)))-1.5});
\draw[domain=-0.74:2, smooth, variable=\x, red] plot ({\x}, {sin(deg((\x/2-1)))-1.5});

\draw[blue] (-0.74,-2.5) -- (-2,-1.24) ;
\draw[blue] (-0.15,-1) -- (-2,0.85) ;

\node at (-0.74,-2.5) [circle, fill, inner sep=2 pt]{};
\node at (-0.5,-2.4) [label=below:$\S_1$]{};

\node at (-0.15,-1) [circle, fill, inner sep=2 pt]{};
\node at (0.08,-0.9) [label=below:$\S_1$]{};

\node at (-1.4,-0.3) [label=below:$\color{red} \Sigma_1$]{};
\node at (1.5,0.4) [label=below:$\color{red} \Sigma$]{};

\node at (-2.1,1) [label=right:$\color{blue} L$]{};

\node at (0.67,-1.35) [circle, fill, inner sep=2 pt, red]{};

\end{tikzpicture} 
\caption{\label{fig:location_contracting_phase}}
    \end{minipage}
\end{figure}
Therefore, letting $\Sigma$ evolve throughout the foliation $\F$, $\S_1$ can follow any timelike (possibly locally lightlike) trajectory in the region corresponding to the intersection of the causal patch of the pode and the region of trapped surfaces, corresponding to the purple triangle in \Fig{fig:location_contracting_phase}. Everything said so far for the holographic screen $\S_1$ associated with an observer at the pode can be adapted for a second screen $\S_2$ associated with an observer at the antipode. In the contracting phase, the trajectory of $\S_2$ is arbitrary in the intersection of the causal patch of the antipode and the region of trapped surfaces, and then has to follow the apparent horizon of the antipode in the expanding phase. The trajectories of $\S_1$ and $\S_2$ might intersect in the region of trapped surfaces, so that $\S_2$ can actually be to the left of $\S_1$. On each slice $\Sigma$ of $\F$, we will then denote $\S_\Le$ the leftmost sphere among $\S_1$ and $\S_2$, and $\S_\Ri$ the rightmost sphere among $\S_1$ and $\S_2$. 

A completely similar analysis can be carried out for the Big Bang/Big Crunch cosmologies with $\gamma\geq 1$; the only difference being that the expanding phase lies in the lower half part of the Penrose diagram, and the contracting phase in the upper half part. Two arbitrary trajectories for the screens $\S_\Le$ and $\S_\Ri$ following from this construction are depicted by the purple and orange curves respectively, both for bouncing and Big Bang/Big Crunch cosmologies, in the Penrose diagrams of \Fig{fig:trajectories}.

\subsection{Bilayer proposal}
\label{sec:bilayer}

\begin{figure}[h!]
\begin{subfigure}[t]{0.48\linewidth}
\centering
\begin{tikzpicture}

\path
       +(2,4)  coordinate (IItopright)
       +(-2,4) coordinate (IItopleft)
       +(2,-4) coordinate (IIbotright)
       +(-2,-4) coordinate(IIbotleft)
      
       ;

\begin{scope}[blend mode=multiply]
\fill[fill=violet!20] (IIbotleft) to [bend right=8] (0,-1.4) to [bend left=25] (0,0) -- (IItopleft) -- cycle;
\fill[fill=orange!20] (IIbotright) to [bend left=8] (0,-1.4) to [bend right=25] (0,0) -- (IItopright) -- cycle;
\end{scope}
       
\draw (IItopleft) --
          node[midway, above, sloped]    {}
      (IItopright) --
          node[midway, above, sloped] {}
      (IIbotright) -- 
          node[midway, below, sloped] {}
      (IIbotleft) --
          node[midway, above , sloped] {}
      (IItopleft) -- cycle;
      
\draw[violet,line width=0.8mm] (IItopleft) -- (0,0);
\draw (0,0) -- (IIbotright);
\draw[orange,line width=0.8mm] (IItopright) -- (0,0);
\draw (0,0) -- (IIbotleft) ;
              
\draw[dashed,gray] (IIbotleft) -- (2/3,-4/3);
\draw[dashed,gray] (IIbotright) -- (-2/3,-4/3);


\draw[domain=-2:2, smooth, variable=\x, red, line width=0.8mm] plot ({\x}, {sin(deg((\x/2-1)))-0.1});

\draw[violet,line width=0.8mm] (IIbotleft) to [bend right=8] (0,-1.4) ;
\draw[violet,line width=0.8mm] (0,-1.4) to [bend left=25] (0,0) ;

\draw[orange,line width=0.8mm] (IIbotright) to [bend left=8] (0,-1.4) ;
\draw[orange,line width=0.8mm] (0,-1.4) to [bend right=25] (0,0) ;

\node at (-0.13,-1) [circle, fill, inner sep=1.5 pt]{};
\node at (-0.35,-0.85) [label=below:$\S_\Le$]{};

\node at (0.15,-0.9) [circle, fill, inner sep=1.5 pt]{};
\node at (0.4,-0.77) [label=below:$\S_\Ri$]{};

\node at (-0.5,0.45) [label=below:$\red \Sigma_\Ex$]{};
\draw[red,-stealth] (-0.5,-0.3) -- (0,-0.85);
\node at (-1.3,-1.2) [label=above:$\red \Sigma_\Le$]{};
\node at (1.3,-0.5) [label=above:$\red \Sigma_\Ri$]{};




\end{tikzpicture} 
\caption{\footnotesize Case of a bouncing cosmology, $\gamma\le -1$.}
\end{subfigure}
\quad \,
\begin{subfigure}[t]{0.48\linewidth}
\centering
\begin{tikzpicture}

\path
       +(2,4)  coordinate (IItopright)
       +(-2,4) coordinate (IItopleft)
       +(2,-4) coordinate (IIbotright)
       +(-2,-4) coordinate(IIbotleft)
      
       ;

\begin{scope}[blend mode=multiply]
\fill[fill=violet!20] (IItopleft) to [bend left=8] (0,1.4) to [bend right=25] (0,0) -- (IIbotleft) -- cycle;
\fill[fill=orange!20] (IItopright) to [bend right=8] (0,1.4) to [bend left=25] (0,0) -- (IIbotright) -- cycle;
\end{scope}
      
\draw[decorate,decoration=snake] (IItopleft) --
          node[midway, above, sloped]    {}
      (IItopright);
      
\draw (IItopright) --
          node[midway, above, sloped] {}
      (IIbotright);
      
\draw[decorate,decoration=snake]  (IIbotright) -- 
          node[midway, below, sloped] {}
      (IIbotleft);
      
\draw (IIbotleft) --
          node[midway, above , sloped] {}
      (IItopleft);
      
\draw[violet,line width=0.8mm] (IIbotleft) -- (0,0);
\draw (0,0) -- (IItopright);
\draw[orange,line width=0.8mm] (IIbotright) -- (0,0);
\draw (0,0) -- (IItopleft) ;
              
\draw[dashed,gray] (IItopleft) -- (2/3,4/3);
\draw[dashed,gray] (IItopright) -- (-2/3,4/3);

\draw[domain=-2:2, smooth, variable=\x, red, line width=0.8mm] plot ({\x}, {sin(deg((\x/2-1)))+3.2});

\draw[violet,line width=0.8mm] (IItopleft) to [bend left=8] (0,1.4) ;
\draw[violet,line width=0.8mm] (0,1.4) to [bend right=25] (0,0) ;

\draw[orange,line width=0.8mm] (IItopright) to [bend right=8] (0,1.4) ;
\draw[orange,line width=0.8mm] (0,1.4) to [bend left=25] (0,0) ;

\node at (-0.5,2.25) [circle, fill, inner sep=1.5 pt]{};
\node at (0.8,2.6) [circle, fill, inner sep=1.5 pt]{};

\node at (-0.6,1.4) [label=above:$\S_\Le$]{};
\node at (0.83,1.75) [label=above:$\S_\Ri$]{};

\node at (0,3.3) [label=below:$\red \Sigma_\Ex$]{};
\node at (-1.5,2.3) [label=below:$\red \Sigma_\Le$]{};
\node at (1.68,3) [label=below:$\red \Sigma_\Ri$]{};

\end{tikzpicture}     
\caption{\footnotesize Case of a Big Bang/Big Crunch cosmology, \mbox{$\gamma\ge 1$}.}
\end{subfigure}
    \caption{\footnotesize Trajectories of the left screen $\S_\Le$ (purple) and the right screen $\S_\Ri$ (orange). The interior region of the pode is purple shaded, while the interior region of the antipode is orange shaded. The exterior region is white. \label{fig:trajectories}}

\end{figure}
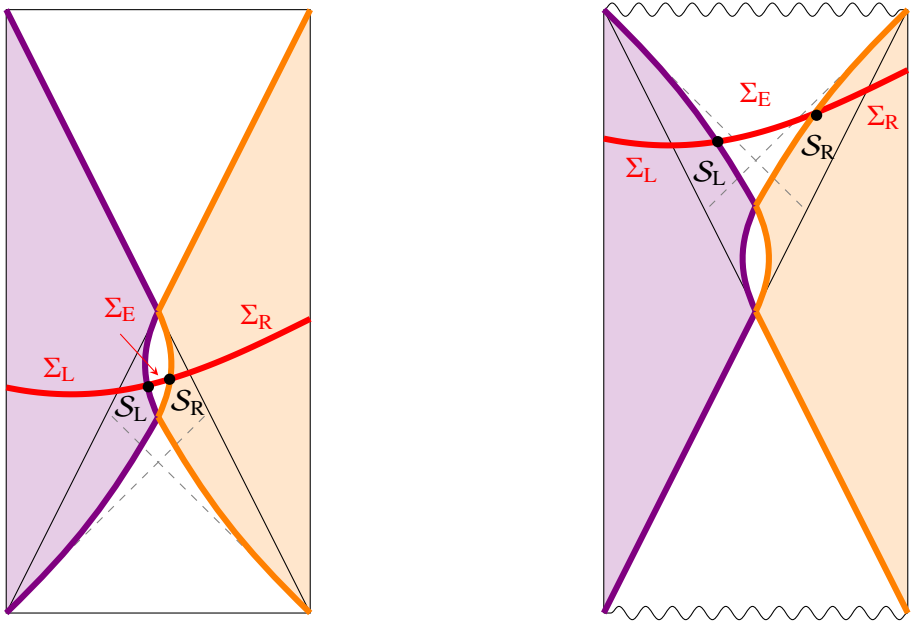

\begin{figure}[h!]
        \centering
	\includegraphics[height=35mm]{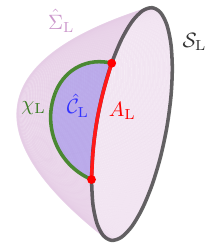}
        \hspace{0.4cm}
	\centering
	\includegraphics[height=35mm]{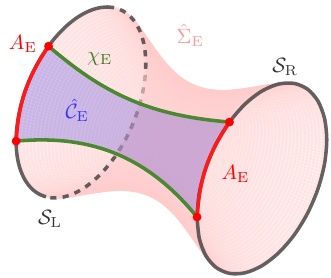}
	\hspace{0.5cm}
	\centering
	\includegraphics[height=35mm]{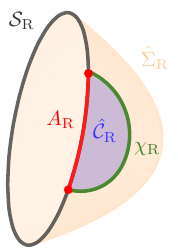}
 \caption{\footnotesize Any Cauchy slice $\hat\Sigma$ passing through the two screens $\S_\Le$ and $\S_\Ri$ is split into three parts: $\hat\Sigma=\hat\Sigma_\Le \cup \hat\Sigma_\Ex \cup \hat\Sigma_\Ri$. Considering a subregion $A$ of $\S_\Le \cup \S_\Ri$, we denote by $A_\Le=A\cap \S_\Le, A_\Ex=A$, and $A_\Ri=A\cap \S_\Ri$, depicted in red. For $i\in\{\Le,\Ex,\Ri\}$, we show in green a codimension-2 surface $\chi_i$ of minimal extremal area, that is homologous to $A_i$ on the Cauchy slice $\hat \Sigma_i$. $\hat \C_i$ is the codimension-1 surface on $\hat \Sigma_i$ bounded by $\chi_i$ and $A_i$, depicted in blue. \label{fig:topology_bilayer}}
\end{figure}

We are now ready to present our bilayer proposal for computing entanglement entropies between complementary subsystems of the pair of screens, as well as the bulk regions that may be reconstructed from such holographic subsystems. For any bouncing or Big Bang/Big Crunch cosmology with $|\gamma|\geq 1$, let us consider a foliation of spacetime with SO$(n)$-symmetric Cauchy slices $\Sigma$, as well as two nowhere spacelike curves following the trajectories described in \Sect{sect:location_screens}. Each Cauchy slice $\Sigma$ intersects these trajectories at two S$^{n-1}$ spheres, where holographic screens $\S_\Le$ (the leftmost) and $\S_\Ri$ (the rightmost) are located. This construction naturally splits $\Sigma$ into three parts: $\Sigma=\Sigma_\Le\cup\Sigma_\Ex\cup\Sigma_\Ri$. $\Sigma_\Le$ ($\Sigma_\Ri$) has the topology of a spherical cap bounded by $\S_\Le$ ($\S_\Ri$), while $\Sigma_\Ex$ has the topology of a cylinder bounded by $\S_\Le \cup \S_\Ri$, see \Figs{fig:trajectories} and~\ref{fig:topology_bilayer}. We refer to the ``interior region of the pode'' the domain spanned by all Cauchy slices $\Sigma_\Le$, and by the ``interior region of the antipode'' the domain spanned by all Cauchy slices $\Sigma_\Ri$. They are respectively shown in purple and orange in \Fig{fig:trajectories}. Finally, the white region in \Fig{fig:trajectories} spanned by all $\Sigma_\Ex$ is called the ``exterior region''. We will label these regions as $\Le$, $\Ri$ and $\Ex$, respectively. Let us note that for any Cauchy slice $\hat\Sigma$ passing through $\S_\Le$ and $\S_\Ri$, $\hat\Sigma_i$ and $\Sigma_i$ have the same causal structure, for $i\in\{\Le, \Ex, \Ri\}$.

\noindent Our bilayer holographic entanglement entropy prescription can now be stated as follows:\\
\noindent $\bullet$ For any subregion $A$ of $\S_\Le\cup \S_\Ri$ on the Cauchy slice $\Sigma$, we denote: 
\begin{equation}
A_\Le=A\cap \S_\Le, \qquad A_\Ex=A, \qquad A_\Ri=A\cap \S_\Ri.
\end{equation}
\noindent $\bullet$ For $i\in\{\Le,\Ex,\Ri\}$, let $\chi_i$ be a codimension-2 surface of minimal extremal area  that is homologous to $A_i$ and lies on a Cauchy slice $\hat \Sigma_i$. At leading order in $G\hbar$, the von Neumann entropy $S$ of the subsystem on $A$, \ie the entanglement entropy between $A$ and its complement in $\S_\Le\cup \S_\Ri$, satisfies 
\begin{equation}\label{eq:classical_vN_entropy}
S(A)={\A(\chi_\Le)+\A(\chi_\Ex)+\A(\chi_\Ri)\over 4G\hbar}+\O((G\hbar)^0),
\end{equation}
where $\A(\chi_i)$ is the area of $\chi_i$.

\noindent $\bullet$ Let $\hat \C_i$ be the codimension-1 surface on $\hat \Sigma_i$ bounded by $\chi_i$ and $A_i$, \ie satisfying $\partial\hat \C_i=\chi_i\cup A_i$. Assuming entanglement wedge reconstruction~\cite{Dong:2016eik,Cotler:2017erl}, the state on $\hat \C_\Le\cup\hat \C_\Ex\cup\hat \C_\Ri$ is dual to the state of the holographic subsystem on $A$. In particular, the von Neumann entropy of the state on $\hat \C_\Le\cup\hat \C_\Ex\cup\hat \C_\Ri$, \ie the entanglement entropy between $\hat \C_\Le\cup\hat \C_\Ex\cup\hat \C_\Ri$ and its complement in $\hat \Sigma$, equals the von Neumann entropy of the holographic subsystem on $A$, \ie the entanglement entropy between $A$ and its complement in $\S_\Le\cup \S_\Ri$. The ``entanglement wedge'' of $A$, which is a spacetime region reconstructible from the dual holographic subsystem on $A$, is the union of the three causal diamonds of $\hat\C_\Le$, $\hat\C_\Ex$, $\hat\C_\Ri$.\\

The Cauchy slice $\hat\Sigma=\hat\Sigma_\Le \cup \hat\Sigma_\Ex \cup \hat\Sigma_\Ri$, together with a subregion $A$ of $\S_\Le\cup\S_\Ri$ and the relevant surfaces $\chi_i$ and $\hat\C_i$, are depicted in \Fig{fig:topology_bilayer}.

In the next section, we apply this bilayer holographic proposal to compute the entanglement entropies associated with the two-screen system $\S_\Le\cup \S_\Ri$ and the single-screen system $\S_\Le$, for cosmologies with $|\gamma|\geq 1$. We will focus on the classical, geometrical contributions to the entropies, which are of order $(G\hbar)^{-1}$. The thermal corrections due to the coarse-grained entropy carried by a thermal gas of particles, which are of order $(G\hbar)^{0}$, can be found in \cite{Franken:2023jas}.

\section{Time-dependent ER=EPR}
\label{sect:time_dep_ER=EPR}
\subsection{The two-screen system in bouncing and Big Bang/Big Crunch cosmologies}
\label{sect:two_screen_system}
\begin{figure}[h!]
        \centering
	\includegraphics[height=35mm]{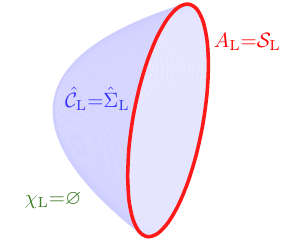}
        \hspace{0.4cm}
	\centering
	\includegraphics[height=35mm]{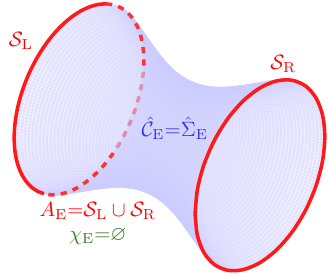}
	\hspace{0.5cm}
	\centering
	\includegraphics[height=35mm]{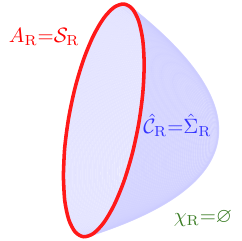}
 \caption{\footnotesize Topology of a Cauchy slice $\hat\Sigma$ passing through the two screens $\S_\Le$ and $\S_\Ri$, and split into three parts: $\hat\Sigma=\hat\Sigma_\Le \cup \hat\Sigma_\Ex \cup \hat\Sigma_\Ri$. For the two-screen system $A=\S_\Le\cup \S_\Ri$ (shown in red), the minimal extremal surface $\chi_i$ homologous to $A_i$ on $\hat \Sigma_i$ is the empty set, while $\hat \C_i=\hat\Sigma_i$ (shown in blue), $\forall i\in\{\Le,\Ex,\Ri\}$. \label{fig:topology_two_screen_system}}
\end{figure}

Let us consider as a first simple example the two-screen system $A=\S_\Le\cup \S_\Ri$ in cosmologies with $|\gamma| \geq 1$. Following the notations introduced in \Sect{sec:bilayer}, we have $A_{\rm L}=\S_\Le$, $A_{\rm E}=\S_\Le \cup \S_\Ri$ and $A_{\rm R}=\S_\Ri$. Let us consider an SO$(n)$-symmetric Cauchy slice $\hat {\Sigma}$ containing $\S_\Le$ and $\S_\Ri$. Since the Cauchy slice $\hat\Sigma_\Le$ ($\hat\Sigma_\Ri$) has the topology of a spherical cap bounded by $A_{\rm L}=\S_\Le$ ($A_{\rm R}=\S_\Ri$), the minimal extremal surface homologous to $A_{\rm L}$ ($A_{\rm R}$) on $\hat\Sigma_\Le$ ($\hat\Sigma_\Ri$) is the empty set, $\chi_\Le=\varnothing$ ($\chi_\Ri=\varnothing$). On the other hand, the Cauchy slice $\hat\Sigma_\Ex$ has the topology of a cylinder bounded by $A_\Ex=\S_\Le\cup\S_\Ri$, so that the minimal extremal surface homologous to $A_\Ex$ on $\hat\Sigma_\Ex$ is again the empty set: $\chi_\Ex=\varnothing$. For $i\in\{\Le,\Ex,\Ri\}$, the homology condition $\partial \hat{\mathcal{C}}_i=\chi_i\cup A_i=\varnothing\cup A_i$ is satisfied for $\hat \C_i=\hat\Sigma_i$. The Cauchy slice $\hat\Sigma$, together with the two-screen system $A=\S_\Le\cup\S_\Ri$ and the relevant surfaces $\chi_i$ and $\hat\C_i$, are depicted in \Fig{fig:topology_two_screen_system}.

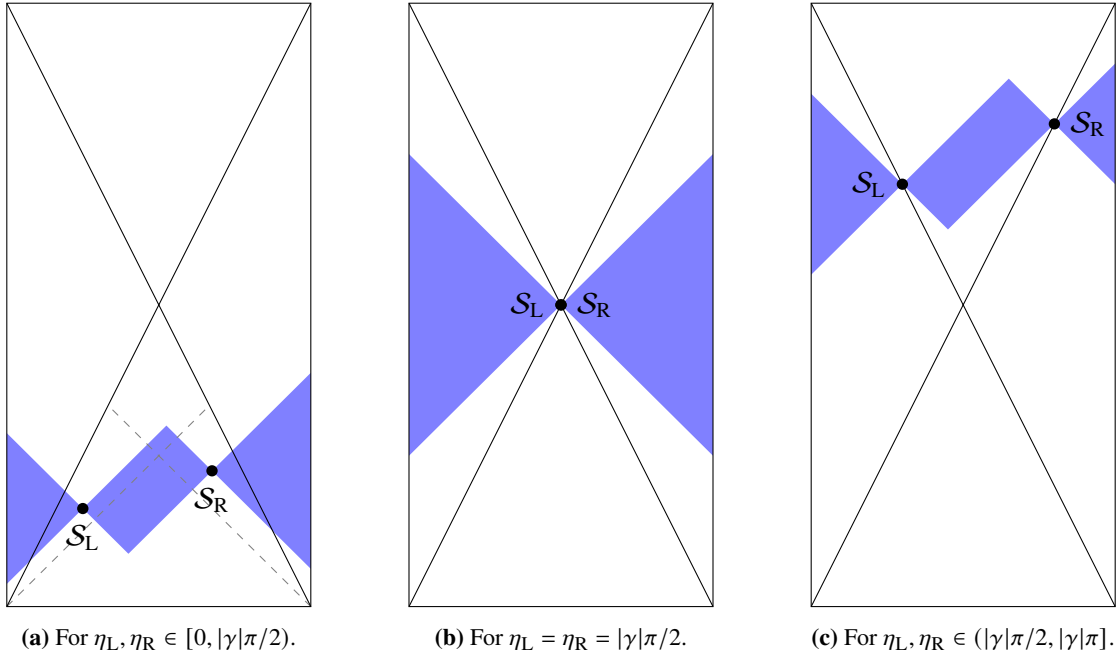
\begin{figure}[h!]
\centering
\begin{subfigure}[c]{0.3\linewidth}
\centering
\begin{tikzpicture}

\path
       +(2,4)  coordinate (IItopright)
       +(-2,4) coordinate (IItopleft)
       +(2,-4) coordinate (IIbotright)
       +(-2,-4) coordinate(IIbotleft)
      
       ;
       
\fill[fill=blue!50] (-1,-2.7) -- (0.1,-1.6) -- (0.7,-2.2) -- (-0.4,-3.3) -- cycle;

\fill[fill=blue!50] (-2,-1.7) -- (-1,-2.7) -- (-2,-3.7) --  cycle;

\fill[fill=blue!50] (2,-0.9) -- (0.7,-2.2) -- (2,-3.5) --  cycle;


\draw[dashed,gray] (IIbotleft) -- (2/3,-4/3);
\draw[dashed,gray] (IIbotright) -- (-2/3,-4/3);
       
\draw (IItopleft) --
      (IItopright) --
      (IIbotright) -- 
      (IIbotleft) --
      (IItopleft) -- cycle;

\draw (IItopleft) -- (IIbotright)
              (IItopright) -- (IIbotleft) ;
      
\node at (-1,-2.7) [circle,fill,inner sep=1.5pt, label = below:$\S_\Le$]{};
\node at (0.7,-2.2) [circle,fill,inner sep=1.5pt, label = below:$\S_\Ri$]{};

\end{tikzpicture}
\caption{\footnotesize For $\eta_\Le, \eta_\Ri\in[0,|\gamma|\pi/2)$.}
\end{subfigure}\hfill
\begin{subfigure}[c]{0.3\linewidth}
\centering
\begin{tikzpicture}

\path
       +(2,4)  coordinate (IItopright)
       +(-2,4) coordinate (IItopleft)
       +(2,-4) coordinate (IIbotright)
       +(-2,-4) coordinate(IIbotleft)
      
       ;
       
\fill[fill=blue!50] (-2,2) -- (0,0) -- (-2,-2) --  cycle;

\fill[fill=blue!50] (2,2) -- (0,0) -- (2,-2) --  cycle;


       
\draw (IItopleft) --
      (IItopright) --
      (IIbotright) -- 
      (IIbotleft) --
      (IItopleft) -- cycle;

\draw (IItopleft) -- (IIbotright)
              (IItopright) -- (IIbotleft) ;
      
\node at (0,0) [circle,fill,inner sep=1.5pt, label = left:$\S_\Le$]{};
\node at (0,0) [circle,fill,inner sep=1.5pt, label = right:$\S_\Ri$]{};

\end{tikzpicture}    
\caption{\footnotesize For $\eta_\Le=\eta_\Ri=|\gamma|\pi/2$.}
\end{subfigure}\hfill
\begin{subfigure}[c]{0.3\linewidth}
\centering
\begin{tikzpicture}

\path
       +(2,4)  coordinate (IItopright)
       +(-2,4) coordinate (IItopleft)
       +(2,-4) coordinate (IIbotright)
       +(-2,-4) coordinate(IIbotleft)
      
       ;
       
\fill[fill=blue!50] (-0.8,1.6) -- (0.6,3) -- (1.2,2.4) -- (-0.2,1) -- cycle;

\fill[fill=blue!50] (-2,2.8) -- (-0.8,1.6) -- (-2,0.4) --  cycle;

\fill[fill=blue!50] (2,3.2) -- (1.2,2.4) -- (2,1.6) --  cycle;


       
\draw (IItopleft) --
      (IItopright) --
      (IIbotright) -- 
      (IIbotleft) --
      (IItopleft) -- cycle;

\draw (IItopleft) -- (IIbotright)
              (IItopright) -- (IIbotleft) ;
      
\node at (-0.8,1.6) [circle,fill,inner sep=1.5pt, label = left:$\S_\Le$]{};
\node at (1.2,2.4) [circle,fill,inner sep=1.5pt, label = right:$\S_\Ri$]{};

\end{tikzpicture}
\caption{\footnotesize For $\eta_\Le, \eta_\Ri\in(|\gamma|\pi/2,|\gamma|\pi]$.}
\end{subfigure}\hfill
     \caption{\footnotesize Entanglement wedge (blue shaded region) of the two-screen system on $\S_\Le \cup\S_\Ri$, when $\gamma \leq -1$. The entire spacetime is covered by the entanglement wedge as the latter evolves during the whole cosmological evolution. \label{fig:Penrose_two_screen_system}}
\end{figure}

The formula \eqref{eq:classical_vN_entropy} therefore gives a vanishing leading classical contribution to the von Neumann entropy of the two-screen system:
\begin{equation}
S(\S_\Le\cup \S_\Ri) =0 +{\mathcal{O}}(G\hbar)^{0}.
\end{equation}
The entanglement wedge of the two-screen system is the union of the causal diamonds of $\hat\Sigma_\Le$, $\hat\Sigma_\Ex$ and $\hat\Sigma_\Ri$, \ie the union of the blue triangles and the rectangle in \Fig{fig:Penrose_two_screen_system}.
In particular, the entanglement wedge comprises slices that are complete Cauchy slices, with respect to the bulk cosmology. Assuming entanglement wedge reconstruction \cite{Dong:2016eik,Cotler:2017erl}, the state on the whole of $\hat \Sigma$ or $\Sigma$ can be reconstructed from the holographic dual system on $\S_\Le\cup \S_\Ri$. Each point of the cosmological spacetime will have been inside the entanglement wedge at least once, as the screens evolve with time. Therefore, we expect to be possible to encode holographically on the two-screen system the state of the cosmology on any slice of any complete foliation $\F$.

\subsection{The single-screen system in bouncing cosmologies}
\label{sect:single_screen_system}
We now consider the single-screen system $A=\S_\Le$ in bouncing cosmologies with $\gamma<-1$. We have in this case $A_{\rm L}=\S_\Le$, $A_{\rm E}=\S_\Le$ and $A_{\rm R}=\varnothing$. The analysis to find the minimal extremal surface homologous to $A_\Le$ in the pode interior region L is the same as for the two-screen system described in \Sect{sect:two_screen_system}: we get $\chi_\Le=\varnothing$ and $\hat\C_\Le=\hat\Sigma_\Le$. In the antipode region R where $A_\Ri=\varnothing$, we trivially get $\chi_\Ri=\varnothing$ and $\hat\C_\Ri=\varnothing$. Therefore, the classical contribution to the von Neumann entropy of the single-screen system arises solely from the exterior region E. Since $A_\Ex=\S_\Le$ has no boundary, any homologous surface on a Cauchy slice $\hat\Sigma_\Ex$ must be closed. Restricting furthermore to SO$(n)$-symmetric surfaces, we are looking for minimal extremal S$^{n-1}$ spheres, represented by points on the Penrose diagram. From the metric \eqref{eq:metric} and the expression of the scale factor \eqref{eq:scale_factor}, one sees that the area of the S$^{n-1}$ sphere located at $(\theta,\eta)$ is given by: 
\begin{equation}
\A(\theta,\eta)=\omega_{n-1}\,a_0^{n-1}\left(\sin{\eta\over |\gamma|}\right)^{\gamma(n-1)}(\sin\theta)^{n-1},
\label{eq:area_function}
\end{equation}
where $\omega_{n-1}$ is the area of the unit S$^{n-1}$.

\begin{figure}[H]
    \begin{minipage}[c]{.55\linewidth}

Since $\chi_\Ex$ must lie on some Cauchy slice $\hat\Sigma_\Ex$ bounded by $\S_\Le\cup\S_\Ri$, one must look for it within the causal diamond of $\Sigma_\Ex$, depicted by the red rectangle in \Fig{fig:min_ext_surfaces_1}. Since this region has a boundary, extremizing $\A$ requires to supplement it with Lagrange multipliers and auxiliary fields enforcing all extremal surfaces to lie in the diamond, including its boundary. Schematically, we are thus looking for extrema of the area function $\hA=\A+\text{Lagrange~multipliers}~\times~\text{constraints}$. If several extremal surfaces exist, $\chi_\Ex$ is the one with the smallest area. As described in details in \cite{Franken:2023jas}, the extremization $\delta\hA/\delta x^{\pm}=0$ then yields four SO$(n)$-symmetric surfaces with extremal areas in the exterior causal diamond, depicted by the colored dots in the Penrose diagram of \Fig{fig:min_ext_surfaces_1}. We thus conclude that the minimal extremal surface $\chi_\Ex$ homologous to $A_\Ex=\S_\Le$ in the exterior region is the sphere sitting at the bottom of the exterior causal diamond, shown by the green dot $M$. The topology of a Cauchy slice $\hat\Sigma$ passing through $\S_L$, $\S_R$ and $M$, together with the single-screen system $A=\S_\Le$ and the relevant surfaces $\chi_i$ and $\hat\C_i$, are depicted in \Fig{fig:topology_single_screen_system}.
        
    \end{minipage}
    \hfill%
    \begin{minipage}[c]{0.4\linewidth}
	\centering
	\begin{tikzpicture}

\path
       +(2,4)  coordinate (IItopright)
       +(-2,4) coordinate (IItopleft)
       +(2,-4) coordinate (IIbotright)
       +(-2,-4) coordinate(IIbotleft)
      
       ;

\fill[fill=red!20] (-3/4,3/2) -- node[pos=0.4, above] {}(0.675,2.925) -- node[pos=0.7, above] {}(1.2,2.4) -- node[pos=0.5, below] {}(-0.225,0.975) -- node[pos=0.3, above] {}(-3/4,3/2);
       
\draw (IItopleft) --
          node[midway, above, sloped]    {}
      (IItopright) --
          node[midway, above, sloped] {}
      (IIbotright) -- 
          node[midway, below, sloped] {}
      (IIbotleft) --
          node[midway, above , sloped] {}
      (IItopleft) -- cycle;
      
\draw (IItopleft) -- (IIbotright)
              (IItopright) -- (IIbotleft) ;

\draw (-2-0.5,-4+0.5) -- (-2,-4);
\draw (-2+0.5,-4+0.5) -- (-2,-4);
\draw (-2-0.5,-4+0.3) -- (-2-0.5,-4+0.5) -- node[midway, left, sloped] {$x^-$} (-2-0.3,-4+0.5) ;
\draw (-2+0.5,-4+0.3) -- (-2+0.5,-4+0.5) -- (-2+0.3,-4+0.5) ;
\node at (-1.6,-4+0.5) [label = right:$x^+$]{};

\draw (-0.2,2.8+0.4) -- (0,3+0.4) -- (0.2,2.8+0.4);

\draw (-3/4,3/2) to [bend right=10] (1.2,2.4) ;
\node at (0.45,2.8) [label=below:$\Sigma_\Ex$]{};

\node at (-3/4,3/2) [circle,fill,inner sep=2pt, orange, label = left:$\S_\Le$]{};
\node at (1.2,2.4) [circle,fill,inner sep=2pt, orange, label = right:$\S_\Ri$]{};
\node at (-0.225,0.975) [circle,fill,inner sep=2pt, OliveGreen!90]{};
\node at (0.675,2.925) [circle,fill,inner sep=2pt, violet]{};
\node at (-0.225,0.975) [circle,fill,inner sep=1.5pt, OliveGreen!90, label = right:$\color{OliveGreen!90} M$]{};

\end{tikzpicture}   
\caption{\footnotesize Causal diamond of a Cauchy slice $\Sigma_\Ex$. The S$^{n-1}$-area function defined in the causal diamond has minima (green), maxima (purple) and saddle points (orange) on the boundary of the diamond. \label{fig:min_ext_surfaces_1}}
    \end{minipage}
\end{figure}
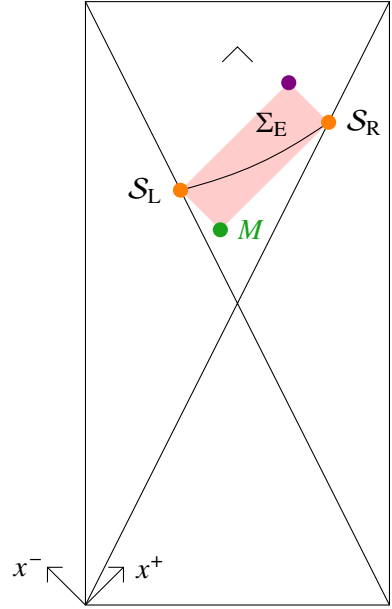

\begin{figure}[h!]
        \centering
	\includegraphics[height=35mm]{left_cap_two_screen.pdf}
        \hspace{0.1cm}
	\centering
	\includegraphics[height=35mm]{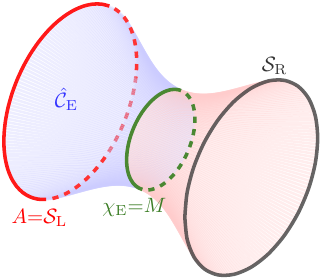}
	\hspace{0.5cm}
	\centering
	\includegraphics[height=35mm]{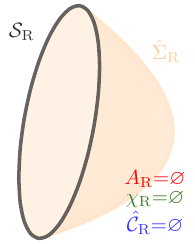}
 \caption{\footnotesize Topology of a Cauchy slice $\hat\Sigma$ passing through the two screens $\S_\Le$ and $\S_\Ri$, and split into three parts: $\hat\Sigma=\hat\Sigma_\Le \cup \hat\Sigma_\Ex \cup \hat\Sigma_\Ri$. Case of the single-screen system $A=\S_\Le$, shown in red. For $i\in\{\Le,\Ex,\Ri\}$, the minimal extremal surfaces $\chi_i$ homologous to $A_i$ on $\hat \Sigma_i$ are shown in green, and the codimension-$1$ surfaces $\hat \C_i$ satisfying the homology constraint $\partial \hat{\mathcal{C}}_i=\chi_i \cup A_i$ correspond to the blue shaded regions.  \label{fig:topology_single_screen_system}}
\end{figure}

The formula \eqref{eq:classical_vN_entropy} yields the leading classical contribution to the von Neumann entropy of the single-screen system:
\begin{equation}\label{eq:entropy_single_screen}
S(\S_\Le) = \frac{\A(M)}{4G\hbar} + {\mathcal{O}}(G\hbar)^{0},
\end{equation}
where $\A(M)$ is the area of the minimal extremal surface $M$ corresponding to the lower vertex of the exterior causal diamond, as depicted in green in \Fig{fig:Penrose_gamma<-1}.
\begin{figure}[h!]
\centering
\begin{subfigure}[t]{0.48\linewidth}
\centering
\begin{tikzpicture}

\path
       +(2,4)  coordinate (IItopright)
       +(-2,4) coordinate (IItopleft)
       +(2,-4) coordinate (IIbotright)
       +(-2,-4) coordinate(IIbotleft)

       +(-3/4,3/2) coordinate (SL)
       +(1.2,2.4) coordinate(SR)
      
       ;

\fill[fill=blue!50] (SL) -- (-2,2.75) -- (-2,0.25) --  cycle;       
\fill[fill=red!20] (-3/4,3/2) -- (0.675,2.925) -- (1.2,2.4) -- (-0.225,0.975) -- (-3/4,3/2);
     
\draw (IItopleft) --
      (IItopright) --
      (IIbotright) -- 
      (IIbotleft) --
      (IItopleft) -- cycle;
      
\draw (IItopleft) -- (IIbotright)
              (IItopright) -- (IIbotleft) ;

\fill[fill=blue!50] (-3/4,3/2+0.05) -- (-0.225,0.975+0.05) -- (-0.225,0.975-0.05) -- (-3/4,3/2-0.05) -- cycle;

\node at (SL) [circle,fill,inner sep=1.5pt,red, label = left:$\color{red} \S_\Le$]{};
\node at (SR) [circle,fill,inner sep=1.5pt, label = right:$\S_\Ri$]{};
\node at (-0.225,0.975) [circle,fill,inner sep=1.5pt, OliveGreen!90, label = right:$\color{OliveGreen!90} M$]{};
\end{tikzpicture}
\caption{\footnotesize At early times, the causal diamond in E is a rectangle with $M$ being the lower vertex. The area of $M$ is non-zero and the classical contribution to the entropy of $\S_\Le$ is non-vanishing. As the screens approach future null infinity, the area of $M$ and the von Neumann entropy of $\S_\Le$ increase.}
\end{subfigure}\hfill
\begin{subfigure}[t]{0.48\linewidth}
\centering
\begin{tikzpicture}

\path
       +(2,4)  coordinate (IItopright)
       +(-2,4) coordinate (IItopleft)
       +(2,-4) coordinate (IIbotright)
       +(-2,-4) coordinate(IIbotleft)
      
       ;
       
\fill[fill=red!20] (-2,4) -- (2,4) -- (0,2) -- cycle;
       
\draw (IItopleft) --
      (IItopright) --
      (IIbotright) -- 
      (IIbotleft) --
      (IItopleft) -- cycle;
      
\draw (IItopleft) -- (IIbotright)
              (IItopright) -- (IIbotleft) ;
              
\fill[fill=blue!50] (-2,4+0.05) -- (0,2+0.05) -- (0,2-0.05) -- (-2,4-0.05) -- cycle;

\node at (IItopleft) [circle,fill,inner sep=1.5pt, red, label = left:$\color{red} \S_\Le$]{};
\node at (IItopright) [circle,fill,inner sep=1.5pt, label = right:$\S_\Ri$]{};
\node at (0,2) [circle,fill,inner sep=1.5pt, OliveGreen!90, label = below:$\color{OliveGreen!90} M$]{};

\end{tikzpicture}
\caption{\footnotesize At large times $\eta_\Le=\eta_\Ri=|\gamma|\pi$, $M$ asymptotes to a fixed point and so the entropy of $\S_\Le$ saturates at a finite value. \label{fig:saturation}}
\end{subfigure}\hfill
    \caption{\footnotesize Penrose diagram for a bouncing cosmology satisfying $\gamma<-1$, when the screens are in the expanding phase. The dark blue region corresponds to the entanglement wedge of the single-screen system $\S_\Le$. The causal diamond in the exterior region~E is red shaded. The minimal extremal sphere, denoted by the green dot $M$, corresponds to the lower vertex.}
    \label{fig:Penrose_gamma<-1}
\end{figure}
The exterior region between the two holographic screens behaves as an Einstein-Rosen bridge, arising from the entanglement between the holographic degrees of freedom on $\S_\Le$ and $\S_\Ri$, according to the ER=EPR conjecture. The Cauchy slice $\hat\Sigma_\Ex$ passing through $\S_L$, $\S_R$ and $M$ acts as an effective bridge, corresponding to the one with the smallest possible bottleneck, where the minimal extremal surface $M$ sits (see \Fig{fig:topology_single_screen_system}). During the cosmological evolution in the expanding phase, the size of this bottleneck increases in time, and so does the entanglement entropy of the single screen-system $S(\S_\Le)$, given in \Eq{eq:entropy_single_screen}. It is from this perspective that we call this realization of the ER=EPR proposal ``time-dependent''.

Denoting $\eta_\Le$ ($\eta_\Ri$) the conformal time of the screen $\S_\Le$ ($\S_\Ri$), the entropy \eqref{eq:entropy_single_screen} increases as $\eta_{\rm L}$ and $\eta_{\rm R}$ increase, and eventually saturates a finite upper bound, as $\eta_{\rm L}, \eta_{\rm R} \to |\gamma|\pi$ (see \Fig{fig:saturation}). This result holds despite the growing to infinity of the area of the screen $\S_\Le$. At later times, the entanglement entropy is a small fraction of the maximal possible value. The extra degrees of freedom added to the screens as these evolve along the apparent horizons remain disentangled. 

This situation has to be contrasted with the de Sitter case $\gamma=-1$, studied in great details in \cite{Franken:2023pni}. In this case, the Penrose diagram is a square, and the apparent horizons coincide with the cosmological horizons. The single-screen system $\S_\Le$ (and $\S_\Ri$) is homologous to a set of degenerate minimal extremal surfaces lying on the part of the cosmological horizons between the two screens. Since the area of the cosmological horizon is constant for $\gamma=-1$, the entropy $S(\S_\Le)$ is constant in time. This de Sitter ER=EPR realization is thus time-independent. The bouncing closed FRW cosmologies $\gamma>1$ can be seen as a classical perturbation of pure de Sitter space, which raises the above mentioned degeneracy and introduces time-dependence in the entanglement entropy between the holographic degrees of freedom.

\subsection{The single-screen system in Big Bang/Big Crunch cosmologies}
Finally, we consider the single-screen system $A=\S_\Le$ in Big Bang/Big Crunch cosmologies with $\gamma\geq 1$. The analysis is completely analogous to the one carried out for the single-screen system in bouncing cosmologies, presented in \Sect{sect:single_screen_system}: we have $A_{\rm L}=\S_\Le$, $A_{\rm E}=\S_\Le$ and $A_{\rm R}=\varnothing$; $\chi_\Le=\varnothing$, $\hat\C_\Le=\hat\Sigma_\Le$ in the pode region L, as well as $\chi_\Ri=\varnothing$ and $\hat\C_\Ri=\varnothing$ in the antipode region R. 

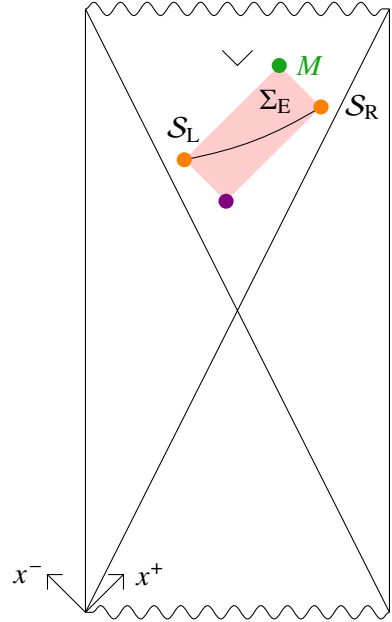
\begin{figure}[H]
    \begin{minipage}[c]{.55\linewidth}

The only non-trivial minimal extremal surface giving a non-vanishing contribution to the von Neumann entropy of the single-screen system is again $\chi_\Ex$. It has to be sought within the causal diamond of $\Sigma_\Ex$, depicted by the red rectangle in \Fig{fig:min_ext_surfaces_2}. This amounts to extremizing the area function \eqref{eq:area_function} supplemented with Lagrange multipliers enforcing all extremal surfaces to lie in the diamond and its boundary. This extremization yields four SO$(n)$-symmetric surfaces \cite{Franken:2023jas} depicted by the colored dots in the Penrose diagram of \Fig{fig:min_ext_surfaces_2}. We thus conclude that the minimal extremal surface $\chi_\Ex$ homologous to $A_\Ex=\S_\Le$ in the exterior region is the sphere sitting at the top of the exterior causal diamond, shown by the green dot $M$. The topology of a Cauchy slice $\hat\Sigma$ passing through $\S_L$, $\S_R$ and $M$ is the same as in the bouncing case $\gamma<-1$, see \Fig{fig:topology_single_screen_system}. The leading classical contribution to the von Neumann entropy of the single-screen system is still given by \Eq{eq:entropy_single_screen}, with $\A(M)$ the area of the minimal extremal surface $M$ corresponding now to the upper vertex of the exterior causal diamond.
        
    \end{minipage}
    \hfill%
    \begin{minipage}[c]{0.4\linewidth}
	\centering
	\begin{tikzpicture}

\path
       +(2,4)  coordinate (IItopright)
       +(-2,4) coordinate (IItopleft)
       +(2,-4) coordinate (IIbotright)
       +(-2,-4) coordinate(IIbotleft)

       +(-0.7,2) coordinate (SL)
       +(1.1,2.7) coordinate (SR)
      
       ;

\fill[fill=red!20] (SL) -- node[pos=0.4, above] {}(0.55,3.25) -- node[pos=0.75, above] {}(SR) -- node[pos=0.5, below] {}(-0.15,1.45) -- node[pos=0.7, below] {}(SL);
       
\draw[decorate,decoration=snake] (IItopleft) --
          node[midway, above, sloped]    {}
      (IItopright);
      
\draw (IItopright) --
          node[midway, above, sloped] {}
      (IIbotright);
      
\draw[decorate,decoration=snake]  (IIbotright) -- 
          node[midway, below, sloped] {}
      (IIbotleft);
      
\draw (IIbotleft) --
          node[midway, above , sloped] {}
      (IItopleft);
      
\draw (IItopleft) -- (IIbotright)
              (IItopright) -- (IIbotleft) ;

\draw (-2-0.5,-4+0.5) -- (-2,-4);
\draw (-2+0.5,-4+0.5) -- (-2,-4);
\draw (-2-0.5,-4+0.3) -- (-2-0.5,-4+0.5) -- node[midway, left, sloped] {$x^-$} (-2-0.3,-4+0.5) ;
\draw (-2+0.5,-4+0.3) -- (-2+0.5,-4+0.5) -- (-2+0.3,-4+0.5) ;
\node at (-1.6,-4+0.5) [label = right:$x^+$]{};

\draw (-0.2,3.2+0.25) -- (0,3+0.25) -- (0.2,3.2+0.25);

\draw (SL) to [bend right=10] (SR) ;
\node at (0.5,3.2) [label=below:$\Sigma_\Ex$]{};

\node at (SL) [circle,fill,inner sep=2pt, orange, label = above:$\S_\Le$]{};
\node at (SR) [circle,fill,inner sep=2pt, orange]{};
\node at (1.15,2.7) [label = right:$\S_\Ri$]{};
\node at (-0.15,1.45) [circle,fill,inner sep=2pt, violet]{};
\node at (0.55,3.25) [circle,fill,inner sep=2pt, OliveGreen!90]{};
\node at (0.55,3.25) [circle,fill,inner sep=1.5pt, OliveGreen!90, label = right:$\color{OliveGreen!90} M$]{};

\end{tikzpicture}   
\caption{\footnotesize Causal diamond of a Cauchy slice $\Sigma_\Ex$. The S$^{n-1}$-area function defined in the causal diamond has minima (green), maxima (purple) and saddle points (orange) on the boundary of the diamond. \label{fig:min_ext_surfaces_2}}
    \end{minipage}
\end{figure}
As described previously, the exterior region between $\S_\Le$ and $\S_\Ri$ behaves as a bridge arising from the entanglement between the holographic degrees of freedom, following the ER=EPR conjecture. The cosmological evolution is shown in \Fig{fig:Penrose_gamma>=1}.
\begin{table}[h!]
\begin{subfigure}[t]{0.3\linewidth}

\centering
\begin{tikzpicture}

\path
       +(2,4)  coordinate (IItopright)
       +(-2,4) coordinate (IItopleft)
       +(2,-4) coordinate (IIbotright)
       +(-2,-4) coordinate(IIbotleft)

       +(-0.7,2) coordinate (SL)
       +(1.1,2.7) coordinate (SR)
      
       ;

\fill[fill=red!20] (SL) -- (0.55,3.25) -- (SR) -- (-0.15,1.45) -- (SL);

\fill[fill=blue!50] (-2,3.3) -- (SL) -- (-2,0.7) --  cycle;

\fill[fill=blue!50] (-0.7,2-0.05) -- (-0.7,2+0.05) -- (0.55,3.25+0.05) -- (0.55,3.25-0.05) -- cycle;

\draw[dashed,gray] (IItopleft) -- (2/3,4/3);
\draw[dashed,gray] (IItopright) -- (-2/3,4/3);

       
\draw[decorate,decoration=snake] (IItopleft) --
          node[midway, above, sloped]    {}
      (IItopright);
      
\draw (IItopright) --
          node[midway, above, sloped] {}
      (IIbotright);
      
\draw[decorate,decoration=snake]  (IIbotright) -- 
          node[midway, below, sloped] {}
      (IIbotleft);
      
\draw (IIbotleft) --
          node[midway, above , sloped] {}
      (IItopleft);
      
\draw (IItopleft) -- (IIbotright)
              (IItopright) -- (IIbotleft) ;


\node at (SL) [circle,fill,inner sep=1.5pt,red]{};
\node at (-0.77,2) [label = above:$\color{red} \S_\Le$]{};
\node at (SR) [circle,fill,inner sep=1.5pt]{};
\node at (1.15,2.7) [label = right:$\S_\Ri$]{};
\node at (0.55,3.25) [circle,fill,inner sep=1.5pt, OliveGreen!90, label = above:$\color{OliveGreen!90} M$]{};
\node at (0,4) [label = above:$\phantom{M}$]{};

\end{tikzpicture}

\end{subfigure}\hfill
\begin{subfigure}[t]{0.3\linewidth}
\centering
\begin{tikzpicture}

\path
       +(2,4)  coordinate (IItopright)
       +(-2,4) coordinate (IItopleft)
       +(2,-4) coordinate (IIbotright)
       +(-2,-4) coordinate(IIbotleft)

       +(-1.1,2.6) coordinate (SL)
       +(1.3,3) coordinate (SR)
      
       ;

\fill[fill=red!20] (SL) -- (0.3,4) -- (SR) -- (-0.1,1.6) -- (SL);

\fill[fill=blue!50] (-2,3.5) -- (SL) -- (-2,1.7) --  cycle;

\fill[fill=blue!50] (-1.1+0.05,2.6) -- (0.3+0.05,4) -- (0.3-0.05,4) -- (-1.1-0.05,2.6) -- cycle;

\draw[dashed,gray] (IItopleft) -- (2/3,4/3);
\draw[dashed,gray] (IItopright) -- (-2/3,4/3);

       
\draw[decorate,decoration=snake] (IItopleft) --
          node[midway, above, sloped]    {}
      (IItopright);
      
\draw (IItopright) --
          node[midway, above, sloped] {}
      (IIbotright);
      
\draw[decorate,decoration=snake]  (IIbotright) -- 
          node[midway, below, sloped] {}
      (IIbotleft);
      
\draw (IIbotleft) --
          node[midway, above , sloped] {}
      (IItopleft);
      
\draw (IItopleft) -- (IIbotright)
              (IItopright) -- (IIbotleft) ;


\node at (SL) [circle,red,fill,inner sep=1.5pt]{};
\node at (-1.17,2.6) [label = above:$\color{red}\S_\Le$]{};
\node at (SR) [circle,fill,inner sep=1.5pt]{};
\node at (1.15,2.7) [label = right:$\S_\Ri$]{};
\node at (0.3,4) [circle,fill,inner sep=1.5pt, OliveGreen!90, label = above:$\color{OliveGreen!90} M$]{};

\end{tikzpicture}           
\end{subfigure}\hfill
\begin{subfigure}[t]{0.3\linewidth}
\centering
\begin{tikzpicture}

\path
       +(2,4)  coordinate (IItopright)
       +(-2,4) coordinate (IItopleft)
       +(2,-4) coordinate (IIbotright)
       +(-2,-4) coordinate(IIbotleft)

       +(-1.2,3) coordinate (SL)
       +(1.5,3.3) coordinate (SR)
      
       ;

\fill[fill=red!20] (SL) -- (-0.2,4) -- (0.8,4) -- (SR) -- (0,1.8) -- (SL);

\fill[fill=blue!50] (-2,3.8) -- (SL) -- (-2,2.2) --  cycle;

\fill[fill=blue!50] (-1.2,3-0.05) -- (-1.2,3+0.05) -- (-0.2,4+0.05) -- (-0.2,4-0.05) -- cycle;

\draw[dashed,gray] (IItopleft) -- (2/3,4/3);
\draw[dashed,gray] (IItopright) -- (-2/3,4/3);


\draw[decorate,decoration=snake] (IItopleft) --
          node[midway, above, sloped]    {}
      (IItopright);
      
\draw (IItopright) --
          node[midway, above, sloped] {}
      (IIbotright);
      
\draw[decorate,decoration=snake]  (IIbotright) -- 
          node[midway, below, sloped] {}
      (IIbotleft);
      
\draw (IIbotleft) --
          node[midway, above , sloped] {}
      (IItopleft);
      
\draw (IItopleft) -- (IIbotright)
              (IItopright) -- (IIbotleft) ;

\node at (SL) [circle,fill,inner sep=1.5pt,red]{};
\node at (-1.27,3) [label = above:$\color{red} \S_\Le$]{};
\node at (SR) [circle,fill,inner sep=1.5pt]{};
\node at (1.75,3.4) [label = below:$\S_\Ri$]{};
\node at (-0.2,4) [circle,fill,inner sep=1.5pt, OliveGreen!90, label = above:$\color{OliveGreen!90} M$]{};

\end{tikzpicture}
\end{subfigure}\\
\begin{subfigure}[t]{0.3\linewidth}
\centering
\includegraphics[height=35mm]{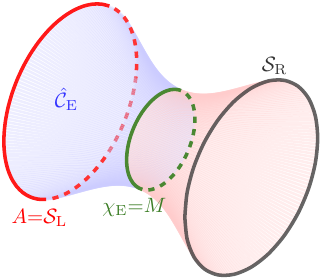}
    \caption{\footnotesize At early times, $M$ has a non-vanishing area, and the effective bridge is smooth and connected.}
\end{subfigure}\hfill
\begin{subfigure}[t]{0.3\linewidth}
\centering
	\includegraphics[height=33mm]{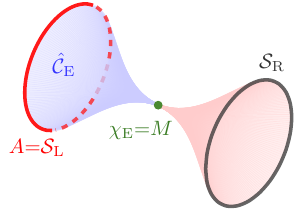}
    \caption{\footnotesize When $M$ hits the Big Crunch singularity, its area vanishes and the effective bridge pinches and closes off. \label{fig:M_hits_singularity_1}}       
\end{subfigure}\hfill
\begin{subfigure}[t]{0.3\linewidth}
\centering
	\includegraphics[height=23mm]{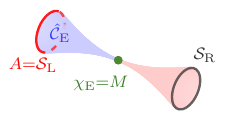}
    \caption{\footnotesize The effective bridge remains disconnected for the rest of the cosmological evolution.}
    \label{fig:M_hits_singularity_2}
\end{subfigure}\hfill
\captionof{figure}{\footnotesize Topology of the effective bridge in the exterior region connecting the screens $\S_\Le$ and $\S_\Ri$ via the minimal extremal surface $M$, depicted at different times in the contracting phase of a Big Bang/Big Crunch cosmology. The Cauchy slice $\hat\Sigma_\Ex$ is the union of the blue and red shaded regions. It corresponds to the limiting Cauchy slice consisting of the two upper edges of the exterior causal rectangle in the Penrose diagram.}
\label{fig:Penrose_gamma>=1}
\end{table}
Contrary to the bouncing case, the entanglement entropy of the single-screen system decreases as $\eta_\Le$ and $\eta_\Ri$ increase, and eventually vanishes when $M$ hits the Big Crunch singularity (see Figures 18b and 18c). Below each Penrose diagrams in \Fig{fig:Penrose_gamma>=1} is represented the Cauchy slice $\hat\Sigma_\Ex$ connecting the screens $\S_\Le$ and $\S_\Ri$ via the minimal extremal surface $M$. When $M$ hits the Big Crunch singularity, the effective bridge pinches and closes off, hence yielding two spatially disconnected slices. This is reminiscent of the qualitative discussion presented in the Introduction, that disentangling the degrees of freedom in the dual theory amounts to destroy spatial connectivity in the gravitational bulk side (see \Fig{fig:intuitive_picture}).

\section{Conclusion}
\label{sect:conclusion}

In this work, we have proposed a covariant holographic description of a generic class of closed FRW cosmologies, arguing that the whole spacetime can be holographically encoded on two screens associated with a pair of antipodal observers. In the expanding phase, the two screens lie at the apparent horizons. In the contracting phase, they can follow an arbitrary timelike trajectory in the overlaps of the causal patches of the observers with the region of trapped spheres. Applying our proposal to compute the entropy of the single-screen subsystem, we find a generic behavior that manifests the ER=EPR conjecture and the connection between quantum entanglement and geometry. Quantum entanglement between the holographic degrees of freedom builds an Einstein-Rosen bridge connecting the two screens, whose size varies in time due to the cosmological evolution. In the expanding phase of the bouncing cosmologies, the entanglement entropy between the two screens increases and eventually saturates an upper bound, despite the fact that the area of the screens grows to infinity. In the contracting phase of the Big Bang/Big Crunch cosmologies, the entanglement entropy decreases and ultimately vanishes. In the geometric picture, this behavior is manifested by the contraction of the effective bridge which eventually pinches and closes off.\\ 

A major breakthrough in the holographic description of closed cosmologies would be to understand the nature of a possible non-gravitational dual quantum theory. In the de Sitter case, it has recently been argued that the high temperature limit of the double-scaled SYK model could provide a dual quantum description of two-dimensional de Sitter JT gravity \cite{Susskind:2021esx,Susskind:2022dfz,Lin:2022nss,Susskind:2022bia,Susskind:2023hnj,Narovlansky:2023lfz,Verlinde:2024znh,Verlinde:2024zrh,Rahman:2022jsf,Rahman:2023pgt,Rahman:2024vyg}. As part of future work, it would be interesting to investigating the nature of a possible holographic dual theory in the closed FRW case as well.

From a different perspective, the importance of observers in closed universes (such as de Sitter or closed FRW cosmologies) has recently been a subject of great interest \cite{Chandrasekaran:2022cip,Witten:2023qsv,Witten:2023xze,Gomez:2023upk,Mirbabayi:2023vgl}. In an asymptotically flat or AdS spacetime, an observer can be located at infinity, hence looking at the spacetime from ``outside'' and having negligible gravitational influence on it. This is not the case in a closed universe, where the observer is necessarily inside the spacetime and does gravitate. In particular, it has been argued that in a closed spacetime, an observer must be included in order to get a non-trivial algebra of observables, generated by the quantum fields along the observer's worldline. It would be particularly interesting to investigate in more details the closed FRW holographic proposal presented in this talk from this point of view. The time-dependent ER=EPR realization observed here is lacking a microscopic explanation, and might be better described and understood from such an algebraic point of view.

\section*{Acknowledgements}
I am very grateful to my collaborators Hervé Partouche, Nicolaos Toumbas, and especially Victor Franken for useful discussions and comments regarding this proceeding. I would like to acknowledge hospitality by the Ecole Polytechnique, where early stages of this work have been done. We are grateful to the organizers of the Corfu Summer Institute 2023 for inviting us to present this talk. This work is partially supported by the Cyprus Research and Innovation Foundation grant EXCELLENCE/0421/0362.

\newpage
\bibliographystyle{jhep}
\bibliography{skeleton}

\end{document}